\documentclass{aa}

\usepackage{graphicx}
\usepackage{txfonts}
\usepackage{caption}
\usepackage{subcaption}
\usepackage{amsmath,amsfonts,bm}
\usepackage{comment}
\begin{document} 

   \title{The operationally ready full three-dimensional magnetohydrodynamic (3D MHD) model from the Sun to Earth: COCONUT+Icarus}

   %\subtitle{?}

   \author{T. Baratashvili \inst{1}, M. Brchnelova\inst{1}, L. Linan\inst{1}, A. Lani\inst{1}, S. Poedts \inst{1,2}
          }

   \institute{Department of Mathematics/Centre for mathematical Plasma Astrophysics, 
             KU Leuven, Celestijnenlaan 200B, 3001 Leuven, Belgium. 
             \email{tinatin.baratashvili@kuleuven.be}
             \and
             Institute of Physics, University of Maria Curie-Sk{\l}odowska, 
             ul.\ Radziszewskiego 10, 20-031 Lublin, Poland}

\date{Accepted: June 28, 2024}
\titlerunning{Full 3D MHD COCONUT corona + Icarus heliosphere model}
\authorrunning{Baratashvili et al.}
% \abstract{}{}{}{}{} 
% 5 {} token are mandatory
 
    \abstract
   {Solar wind modelling has become a crucial area of study due to the increased dependence of modern society on technology, navigation, and power systems. Accurate space weather forecasts can predict upcoming threats to Earth's geospace and allow for harmful socioeconomic impacts to be mitigated. Coronal and heliospheric models must be as realistic as possible to achieve successful predictions. In this study, we examine a novel full magnetohydrodynamic (MHD) chain from the Sun to Earth.}  %leave it empty if necessary  
  % aims headng (mandatory)
   {The goal of this study is to demonstrate the capabilities of the full MHD modelling chain from the Sun to Earth by finalising the implementation of the full MHD coronal model into the COolfluid COroNa
UnsTructured (COCONUT) model and coupling it to the MHD heliospheric model Icarus. The resulting coronal model has significant advantages compared to the pre-existing polytropic alternative, as it includes more physics and allows for a more realistic modelling of bi-modal wind, which is crucial for heliospheric studies. In particular, we examine different empirical formulations for the heating terms in the MHD equations to determine an optimal one that would be able to mimic a realistic solar wind configuration most accurately.}
   {New heating source terms were implemented into the MHD equations of the pre-existing polytropic COCONUT model. A realistic specific heat ratio was applied. In this study, only thermal conduction, radiative losses, and approximated coronal heating function were considered in the energy equation. Multiple approximated heating profiles were examined to see the effect on the solar wind. The output of the coronal model was used to onset the 3D MHD heliospheric model Icarus. A minimum solar activity case was chosen as the first test case for the full MHD model. The numerically simulated data in the corona and the heliosphere were compared to observational products. First, we compared the density data to the available tomography data near the Sun and then the modelled solar wind time series in Icarus was compared to OMNI 1-min data at 1~AU.}
  % Results
   {A range of approximated heating profiles were used in the full MHD coronal model to obtain a realistic solar wind configuration. The bi-modal solar wind was obtained for the corona when introducing heating that is dependent upon the magnetic field. The modelled density profiles are in agreement with the tomography data. The modelled wind in the heliosphere is in reasonable agreement with observations. Overall, the density is overestimated, whereas the speed at 1~AU is more similar to OMNI 1-min data. The general profile of the magnetic field components is modelled well, but its magnitude is underestimated.   }
   {We present a first attempt to obtain the full MHD chain from the Sun to Earth with COCONUT and Icarus. The coronal model has been upgraded to a full MHD model for a realistic bi-modal solar wind configuration. The approximated heating functions have modelled the wind reasonably well, but simple approximations are not enough to obtain a realistic density-speed balance or realistic features in the low corona and farther, near the outer boundary. The full MHD model was computed in 1.06 h on 180 cores of the Genius cluster of the Vlaams Supercomputing Center, which is only 1.8 times longer than the polytropic simulation. The extended model gives the opportunity to experiment with different heating formulations and improves the approximated function to model the real solar wind more accurately.
   }

   \keywords{}

   \maketitle
%
%-------------------------------------------------------------------

\section{Introduction} \label{section:introduction}
Space weather has become a prevailing branch of physics, aimed at studying the environment around the Sun, with a major focus near Earth. Charged particles are continuously emitted from the Sun, following the magnetic field lines of the Sun, forming the so-called solar wind. The latter can interact with the magnetic field of Earth. Moreover, frequently, various eruptions such as coronal mass ejections (CMEs) happen on the solar surface. These CMEs are magnetised clouds that range between $10^{13}\;$g up to $10^{16}\;$g in mass and erupt with velocities ranging from $\sim 100\;\text{km s}^{-1}$ to $\sim 3 000\;\text{km s}^{-1}$ (based on SOHO/LASCO measurements). Thus, they end up carrying enormous momentum and often causing strong shocks during their evolution in the inner heliosphere. The CMEs, when directed towards Earth, can cause significant damages, interfere with spaceborne operations and telecommunication and navigation systems, and provoke severe power outages. Similar damages can also be caused on Earth upon the interaction with the co-rotating interaction region (CIR) shocks \citep{Alves2006}. CMEs usually erupt only a few times during the solar minima, but during solar maxima, a few CMEs can occur during the day \citep{Park2012}. The damages coming from Earth-directed strong CMEs can be partially mitigated only if they are predicted sufficiently in advance. Space weather forecasting has become a standard approach to modelling the solar heliosphere and propagating CMEs in order to avoid disastrous scenarios on Earth. This requires the whole region from the Sun to Earth to be modelled. The dominant physics phenomena along the Sun-Earth path tend to vary. Modelling the whole domain with a single tool is certainly possible \citep{regnault2023}, but since the implemented conditions strongly affect the time step in the simulation and vary from the solar corona to the heliosphere, separating them into two modelling tools is preferable. Moreover, since the domain is large and many grid points are necessary to cover the whole domain, the overall slowdown is significant regarding the simulation wall-clock times. The common practice is to model the solar corona and heliosphere separately based on the physical conditions, since beyond the Alfvén point, all the information travels radially outwards and nothing goes back to the Sun. The natural separation point between the coronal and heliospheric models is the Alfvén point, which is conventionally assumed to be at 0.1~AU. Currently, a number of such forecasting tools exist. Examples of physics-based forecasting tools include ENLIL \citep{Odstrcil2004}, EUHFORIA \citep{Pomoell2018}, SUSANOO-CME \citep{Shiota2016}, and CORHEL \citep{riley2012}, as well as one of the most advanced magnetohydrodynamic (MHD) coronal model called MAS (Magnetohydrodynamic Algorithm outside a Sphere; \citealt{MikicLinker1996,Mikic2018, downs2013}), AWSoM \citep{Vanderholst2014}, and WindPredict \citep{reville2016, reville2020}. These tools model the solar wind with superposed CMEs, propagating them to the Earth and beyond. Currently, ENLIL and EUHFORIA are used in operational settings, performing daily simulation runs to model the solar wind configuration and the CME evolution. The MAS and AWSoM codes have an advanced MHD corona, including a wave turbulence model introduced by Alfvén waves, as well as the transition region in the simulations. Such complex numerical MHD models usually require considerable CPU time to provide a suitable environment for testing different physics mechanisms in investigations of the coronal heating problem, the bi-modal nature of the wind, evolution of complex flux rope models, and the acceleration of solar energetic particle (SEP) events  \citep{RoussevSokolov2006}. 

Recently, a novel 3D MHD polytropic coronal model COCONUT (COolfluid COroNa UnsTructured) was developed at the KU Leuven Centre for mathematical Plasma Astrophysics (\cite{Perri2022}, \cite{Brchnelova2022}). COCONUT has been implemented within the COOLFluiD (Computational Object-Oriented Libraries for Fluid Dynamics) architecture (\cite{Lani2005}, \cite{Kimpe2005}, \cite{Lani2013}, \cite{Lani2014}). COCONUT was developed to become the main MHD coronal model, along with the semi-empirical Wang-Sheeley-Arge (WSA) model \citep{Arge2004}, \citep{Wang1990}, serving for forecasting purposes within the EUropean Heliospheric FORecasting Information Asset (EUHFORIA) 2.0 project \citep{Pomoell2018}. Thus, its domain extends from the solar surface to 0.1~AU. Various studies have been performed with COCONUT so far. In particular, the effects of pre-processing of the magnetograms, which are used for the inner boundary conditions, has been investigated in \cite{Kuzma2023}, while the influence of the magnetogram types on solar corona simulation results was examined in \cite{Perri2022a}. In the present study, the results with a full MHD COCONUT model using a realistic adiabatic index ($\gamma = 5/3$) are demonstrated. New source terms are implemented into the MHD energy equation, representing the contribution from the radiative losses, thermal conduction and coronal heating. In the first approach, the coronal heating is approximated with different functions and the results are analysed. The output of COCONUT is used as the inner boundary condition for Icarus, a recently developed heliospheric modelling tool (\cite{Verbeke2022}, \cite{Baratashvili2022}). Icarus is a 3D MHD heliospheric model developed within the framework of MPI-AMRVAC \citep{Xia2018} and its domain covers radial distances from 0.1~AU to 2~AU, where the solar wind is modelled for the given date and CMEs can be injected in the simulation from the inner boundary. COCONUT and Icarus represent the first fully 3D MHD forecasting chain starting from the Sun to Earth within the EUHFORIA project. The produced solar wind profiles by COCONUT+Icarus chain are then validated by comparing to the observational data. The comparison is extensive and focusses on the effects and profiles of the solar wind both in the low corona, close to the Sun, and farther in the domain, near the Earth. Therefore, the profiles obtained by COCONUT are compared to tomography data, white light images, and OMNI data at 1~AU to confirm their results. 

The paper is organised as follows. Section \ref{coconut_model_description} describes the standard COCONUT numerical model, including the optimisation techniques to make such computationally expensive simulations feasible for forecasting purposes.  Section~\ref{source_terms} introduces the full MHD coronal model and discusses the additional source terms in detail. Afterwards, the obtained results with different approximated heating profiles are discussed with a comparison to the observational data in Sect.~\ref{source_terms}. Section~\ref{coupling_icarus} introduces the Icarus model and presents a discussion of the coupling between COCONUT and Icarus. Finally, our conclusions and outlook are presented in Sect.~\ref{conclusions}.

\section{COCONUT Model} \label{coconut_model_description}
\begin{figure*}[htb!]
    \centering
    \includegraphics[width=0.48\textwidth]{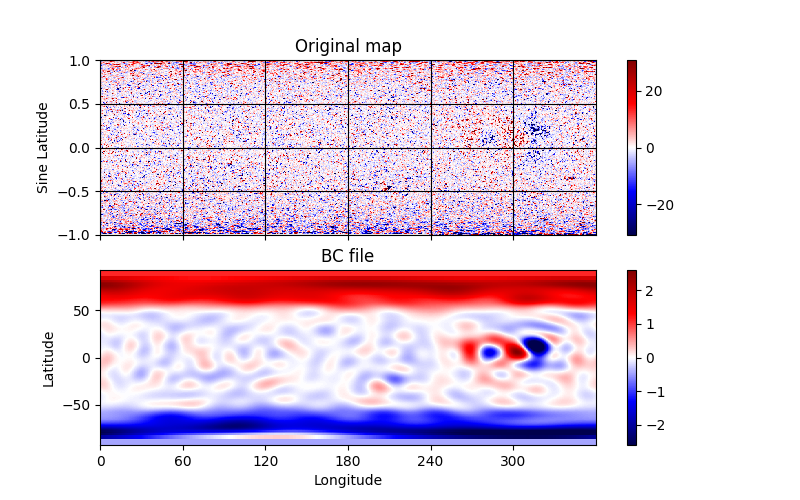}
    \includegraphics[width=0.48\textwidth]{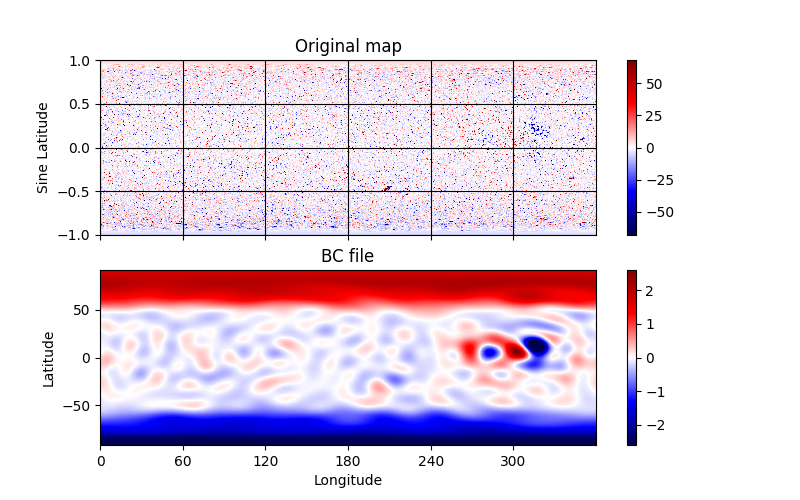}
    \caption{Synoptic maps for July 2, 2019, Carrington Rotation 2219. The original (left) and pole filled (right) HMI magnetograms are provided in the upper panel. The bottom panel shows the processed corresponding input magnetogram with spherical harmonics decomposition filtered with $l_{max} = 20$. The units are in Gauss. In the processed magnetogram, the values are given in the code values (divided by 2.2 G).}
    \label{fig:hmi_original}
\end{figure*}
COCONUT is a 3D MHD model developed as a polytropic model in its first phase. As it was developed potentially as a part of the forecasting chain, the code is highly optimised to produce results fast enough to forecast space weather. To this end, COCONUT was developed as an implicit finite volume (FV) solver, enabling steady-state solutions up to 30 times faster than explicit solvers, as the Courant–Friedrichs–Lewy (CFL) numbers are not limited to 1, but can go up to 10,000 depending on the complexity of the magnetic field configuration \citep{Perri2022}. The use of unstructured grids also contributes to speeding up convergence since it allows for avoiding singularities near the poles, where local time steps tend to get very small and spurious fluxes might be generated.

In COCONUT, the ideal MHD equations are solved in conservative form in Cartesian coordinates, as described in detail in \cite{Perri2022}:

\begin{multline}
\frac{\partial}{\partial t}\left(\begin{array}{c}
\rho \\
\rho \vec{V} \\
\vec{B} \\
E \\
\phi
\end{array}\right)+\vec{\nabla} \cdot \left(\begin{array}{c}
\rho \vec{V} \\
\rho \vec{V} \vec{V}+\tens I\left(p+\frac{1}{2}|\vec{B}|^{2}\right)-\vec{B} \vec{B} \\
\vec{V} \vec{B}-\vec{B} \vec{V}+\underline{\tens I \phi} \\
\left(E+p+\frac{1}{2}|\vec{B}|^{2}\right) \vec{V}-\vec{B}(\vec{V} \cdot \vec{B}) \\
V^2_\text{ref}\mathbf{B}
\end{array}\right) \\ =\left(\begin{array}{c}
0 \\
\rho \vec{g}\\
0 \\
\rho \vec{g} \cdot \vec{V} + \mathbf{S} \\
0
\end{array}\right),
\end{multline}
where ${E}$ is the total energy $\rho \frac{V^2}{2} + \rho \mathcal E + \frac{B^2}{\mathbf{2}}$, $\vec{B}$ is the magnetic field, $\vec{V}$ the velocity, $\vec{g}$ the gravitational acceleration, $\rho$ the density, and $p$ is the thermal gas pressure. The gravitational acceleration is given by $\vec{g}(r) = -(G M_\odot/r^2)\, \hat{\vec{e}}_r$ and the identity dyadic $ \tens I = \hat{\vec{e}}_x \otimes \hat{\vec{e}}_x + \hat{\vec{e}}_y \otimes \hat{\vec{e}}_y + \hat{\vec{e}}_z \otimes \hat{\vec{e}}_z$. The closure is given by the ideal equation of state, thus giving for the internal energy density $\rho \mathcal E = P/(\gamma - 1),$ with a reduced adiabatic index of 1.05. The solar rotation is considered by prescribing a $V_\phi$ component at the inner boundary \citep{Kuzma2023}. Then, \textbf{S} in the energy equation stands for the source terms, which we introduce and explain in detail in Sect. \ref{source_terms}. In the polytropic COCONUT simulations, we have \textbf{S}=0. 

We used an unstructured sixth-level subdivided geodesic polyhedron mesh extended radially outwards in layers between $r = R_\odot$ (inner boundary) and $r = 30.0  \ R_\odot$ (outer boundary), as extensively explained in \cite{Brchnelova2022b}. The mesh consists of 1.9 million cells with increasing cell size in the radial direction. Unlike in the original description of COCONUT in \cite{Perri2022}, in this study, we used more consistent boundary conditions to reduce the generation of the artificial electric field, as described in \citep{Brchnelova2022}.

The typical parameters are prescribed at the solar surface, namely $\rho_\odot = 1.67 \times 10^{-16}\ \mathrm{g/cm^3}$ and similarly for the temperature, $T_\odot = 1.5 \times 10^6\ \mathrm{K}$, are used for fixed-value Dirichlet conditions of density and pressure. Then, from the ideal gas law, the surface value of the pressure can be obtained: $P_\odot = 4.15 \times 10^{-2} \, \mathrm{dyn/cm^2}$ with the following formula:
\begin{equation}
    P_\odot = \frac{2\rho_\odot k_B T_\odot}{\mu m_H},
\end{equation}
where $\rho_\odot$ is the solar surface density, $k_B$ is the Boltzmann constant, $T_\odot$ is the solar surface temperature, $\mu$ is the molecular weight and $m_H$ is the mass of hydrogen. In the original approach, $\mu \sim 1.27$ \citep{Aschwanden2005}. According to \cite{Aschwanden2005}, the total mass density in the fully ionised gas consists of electron and ion densities. The contribution from the most abundant elements, hydrogen and helium, is also considered. However, in \citep{Aschwanden2005}, the number densities of ions are considered to be equal; whereas, in reality, they have different masses and charges, violating the $n_e \sim n_i$ assumption. When taking into consideration that $m_{He} = 4* m_H$, we see that $\mu \sim 1.27$ corresponds to Helium abundance of $\sim 18 \%$, which is a strong overestimation for the solar corona, especially since (in the most recent solar cycles) the Helium abundance is observed to be of the order of $1-2\%$ (\cite{yogesh2021}, \cite{moses2020}). Since the helium abundance is observed to be very small, in this approach, we fixed $\mu =1$, which corresponds to 0 helium abundance in the solar corona. 
 
A Dirichlet boundary condition was prescribed for the radial component of the magnetic field $B_r$, which is directly derived from the magnetic map. The magnetograms represent the magnetic field configuration in the photosphere, where the field is much stronger than at the base of the corona. The magnetogram is pre-processed to smooth the magnetic field, which serves as the inner boundary condition in COCONUT. As suggested in \cite{Kuzma2023}, we used spherical harmonics decomposition to filter the high spherical harmonics beyond a certain $l_{max}$ value. This approach decreases the resolution of the magnetic field features and their strength.

Since the currently employed magnetograms provide only the information about the radial magnetic field component, we allowed the magnetic field in the other directions ($B_\theta$ and $B_\phi$) to evolve freely throughout the simulation. We have seen that this approach leads to an accurate placement of electromagnetic features in the domain; as, for example, in the work of \cite{Kuzma2023}. We also plan to experiment with vector magnetograms that would constrain all three magnetic field components instead. The velocity field at the inner boundary was set so the plasma follows the magnetic field lines, as described in \cite{Brchnelova2022}.

\subsection{Choice of magnetogram} \label{magnetogram_choice}
The input magnetogram can strongly influence the features of the obtained solar wind in the corona. The effect of using different magnetograms was studied by \cite{Perri2022a}. As a conclusion of this extensive study, the simulations based on Helioseismic and Magnetic Imager (HMI) produce the best results. Thus, in this study, only HMI magnetograms were used. However, we further looked into HMI products and decided to use the product that uses interpolations near the poles, since these are the regions with the fewest (and worst) observations. This product provides interpolated south and north poles of the Sun instead of data with significant noise.

The left figure in Fig.~\ref{fig:hmi_original} shows the original HMI product choice. The upper panel shows the HMI magnetogram, while the bottom panel shows the processed input magnetogram for COCONUT at the solar surface. The processing is done as in the original description in \cite{Perri2022}. $l_{max} = 20$ is used for spherical harmonic decomposition. The units are given in Gauss, while in the processed image, the values are code units, meaning the magnetic field is divided by 2.2~G. Positive values can be noticed near the south pole in the processed magnetogram, while that trend is not observed in the original magnetogram. Also, when looking at the original magnetogram, strong noise is present near the poles.

The right figure in Fig.~\ref{fig:hmi_original} represents the pole interpolated HMI magnetogram product. Here, the strength of the magnetic field is higher because it is a higher-resolution map than in the original HMI magnetogram case. After processing, we see that the strength of the obtained input boundary condition file is similar to the original one. The main features in the mid-latitudes also remain unaffected, where we see notable differences near the poles. The poles here are more homogeneous, without any unexpected positive values in the South Pole.
 
In comparing the two products, the pole interpolated HMI product in Fig.~\ref{fig:hmi_original} was chosen for this study to avoid artificially generated artefacts in the modelled corona.

\subsection{Polytropic corona for CR 2219}
\begin{figure}[htp]
    \centering
    \includegraphics[width=0.45\textwidth]{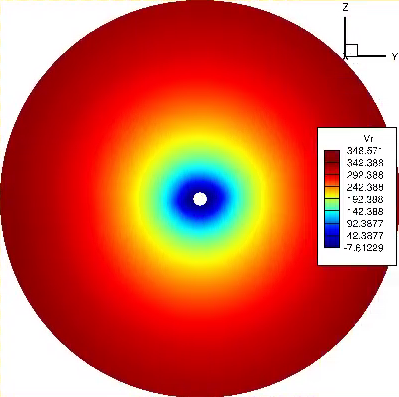}
    \caption{Polytropic solution for the CR2219 corresponding to July 2, 2019. The radial velocity is plotted in the meridional plane. }
    \label{fig:cr2219_polytropic_vr}
\end{figure}

First, we present results for the chosen Carrington Rotation with the polytropic COCONUT model, which was the standard COCONUT model used in all previous works. In this simulation, we used the magnetogram shown in the right panel of Fig.~\ref{fig:hmi_original} as the initial magnetic field configuration at the solar surface. Figure~\ref{fig:cr2219_polytropic_vr} shows the obtained solution for the radial velocity configuration in the meridional plane. The speed distribution is uniform near the outer boundary, namely, $\sim 0.1$ AU with 350~km s$^{-1}$ values.

\begin{figure}[htb!]
    \centering
    \includegraphics[width=0.5\textwidth]{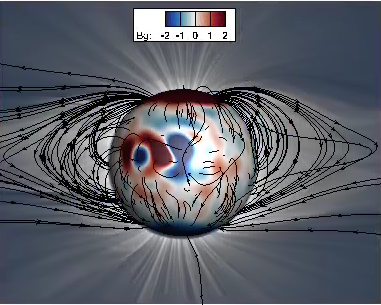}
    \caption{Polytropic solution for the CR2219 corresponding to July 2, 2019. The eclipse image is overlaid with the magnetic streamlines from the simulation. The surface of the Sun is coloured with the $B_r$ in Gauss. }
    \label{fig:cr2219_polytropic_fieldlines}
\end{figure}

Figure~\ref{fig:cr2219_polytropic_fieldlines} shows the same plane as the previous figure but zoomed in on the low corona. The eclipse image is plotted in the background and the magnetic field lines are overlaid by the polytropic COCONUT simulation. The solar surface is coloured with the radial magnetic field values. The large structures seen in the low corona are well mimicked by the magnetic streamlines. This is expected, since plasma is believed to be trapped by the closed loops, enhancing the density in these areas \citep{Lionello2001}.

\begin{figure*}[hbt!]
    \centering
    \includegraphics[width=0.48\textwidth]{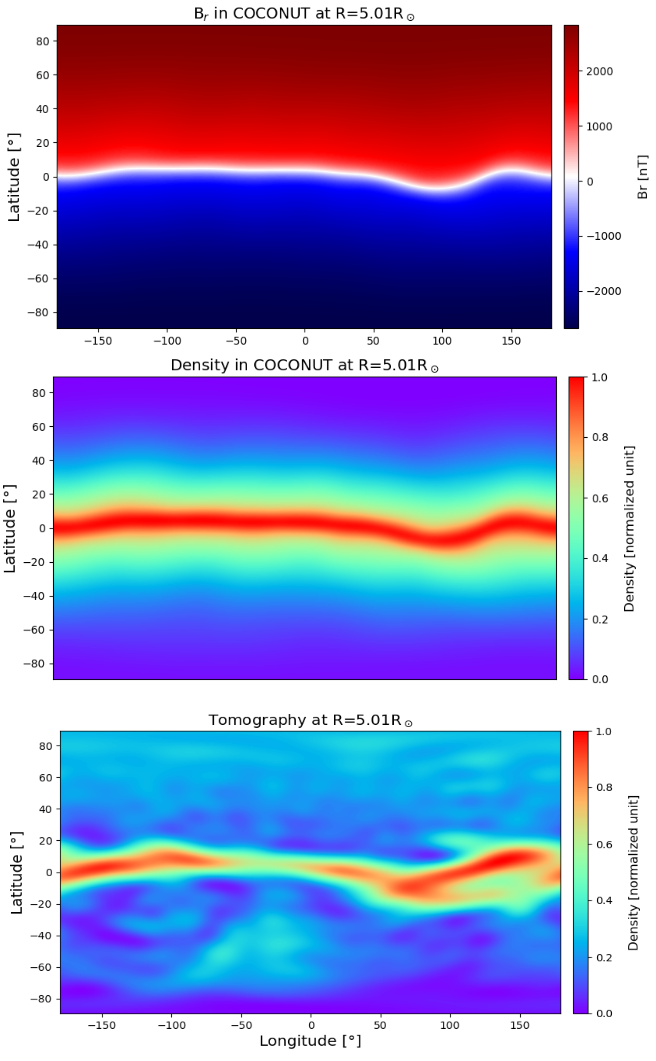}
    \includegraphics[width=0.48\textwidth]{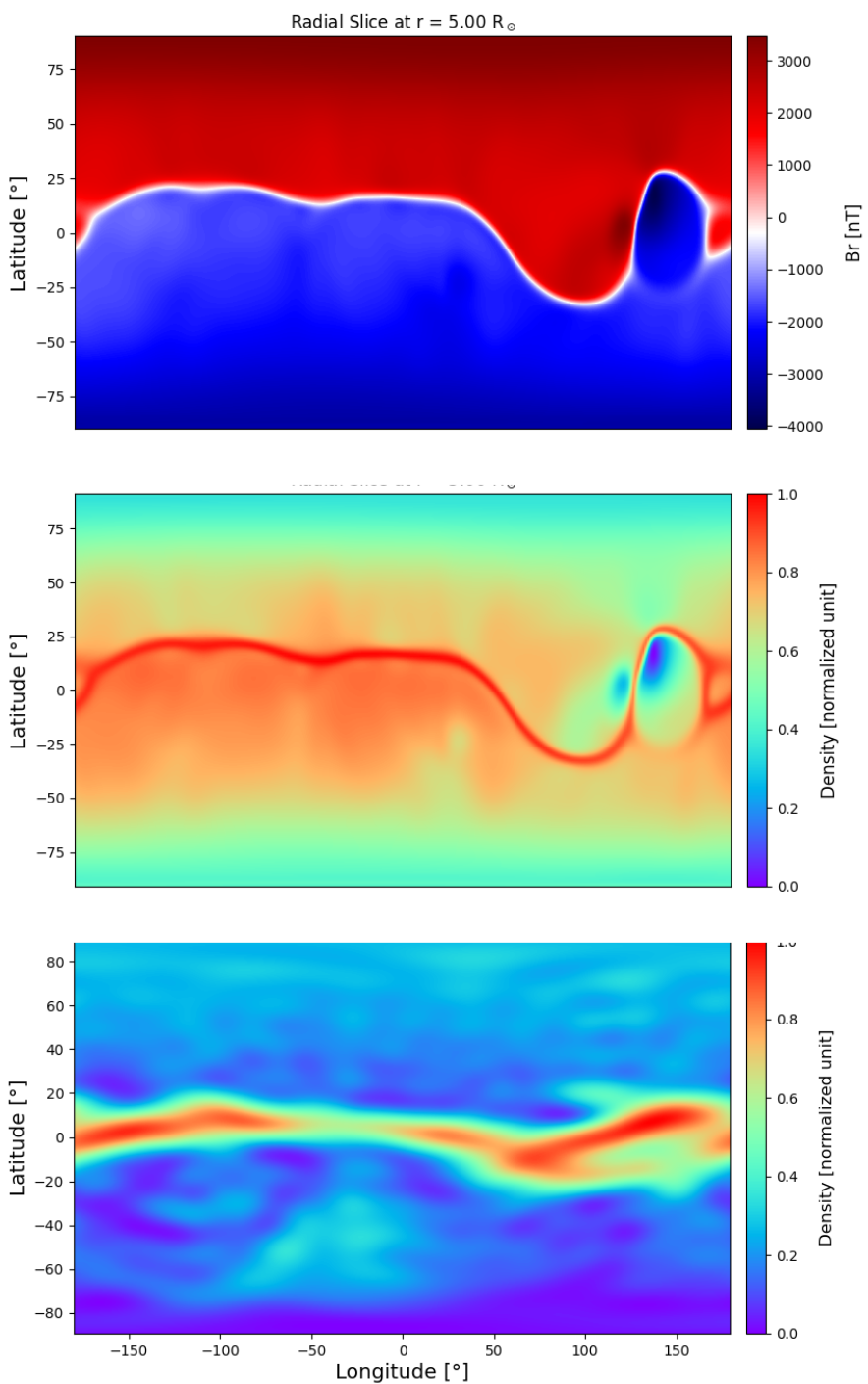}
    \caption{Horizontal axis shows longitudes in degrees and the vertical axis shows co-latitudes in degrees. The left and right panels show the  $B_r$ and density in normalised units from the COCONUT and MAS models, respectively. The bottom row shows the tomography in normalised units from top to bottom, respectively. All quantities are plotted at 5 R$_\odot$. Data from COCONUT are taken from the polytropic simulation. }
    \label{fig:tomography_polytropic}
\end{figure*}

Furthermore, we also compared the modelled data to available tomography data. Coronal rotational tomography is used to estimate the electron density in the coronal plasma (\cite{Morgan2015}, \cite{Morgan2019}, and \cite{Morgan2020}). The available data is based on the STEREO-A COR2 coronagraph polarised brightness observations. The density profiles are given from 4 to 8 solar radii. When comparing the results from COCONUT simulations to the topography data, the comparison to the similar simulation results from the MAS model (\cite{MikicLinker1996}, \cite{Lionello2014}) was also considered.  MAS is a time-dependent resistive thermodynamic MHD model, where the MHD equations are solved on a non-uniform, logically rectangular staggered grid using finite differences. MAS simulation results are available freely through the website \url{www.predsci.com}. Therefore, we chose the data corresponding to the magnetogram for the 2019 eclipse and visualised the results for the polytropic MAS model from the website. 
\\
Figure~\ref{fig:tomography_polytropic} shows the results from the COCONUT simulation (left) and the MAS model (right). The first two rows show the radial magnetic field and number density at 5 R$_\odot$ in the polytropic COCONUT and MAS simulations. The third panel shows the identical tomography reconstruction at 5 R$_\odot$ from the observed data. The density is given in normalised units to better emphasise  the structures obtained at 5 R$_\odot$. This figure shows that the density profile closely resembles the current sheet in the COCONUT results. This is expected since HCSs are characterised by the slow and high-density solar wind \citep{reville2023,Poirier2021}. The density enhancement position agrees with the tomography data; however, the detailed structure is missing. We spot the same behaviour in the MAS simulation results, where the data is in higher resolution, and the current sheet has a different shape. However, the density enhancements and the current sheet are also similar in the MAS model.  

MAS and COCONUT are different codes with different numerical setups, namely, using different grids and numerical methods and boundary conditions \citep{Lionello1999}, as mentioned earlier in this paper. This comparison aims to verify whether these two different solvers would produce similar results. When comparing the features present in the radial magnetic field and, consequently, in the number density plots, we see the enhancements and features present at the same places. The current sheet has a different shape in the MAS simulation, which is partially because the processing of the HMI magnetogram is different and their mesh is much finer, which leads to less numerical dissipation, together with the other above-mentioned differences. 

\section{Full MHD coronal model} \label{source_terms}
The real processes happening near the solar surface that propagate towards the Earth are very complex and not fully understood. Since our goal is to include more physics phenomena occurring near the solar surface, we implemented and tested some empirical source terms that are aimed at approximating the physics of the solar corona more closely. To this end, we followed the paper by \cite{Mikic1999}, \cite{Lionello2009}, and \cite{Downs2010}. In the upgraded MHD model, we fixed the adiabatic index to $\gamma = 5/3$ and used the same inner boundary conditions as in the polytropic COCONUT simulations. Such source terms are introduced in the energy equation, according to the following formulation: 
\begin{equation}
    S = - \nabla \cdot \mathbf{q} + Q_{rad} + Q_{H},
\end{equation}
where $Q_{H}$ is the coronal heating, $Q_{rad}$ is the radiation loss function \cite{Mikic1999} and $- \nabla \cdot \mathbf{q}$ models the thermal conduction. In this study, we neglect resistivity and viscosity and do not introduce these terms in our MHD equations, as their contribution must be small, and we focus on the dominant terms first. 
\begin{figure}[hbt!]
    \centering
    \includegraphics[width=0.5\textwidth]{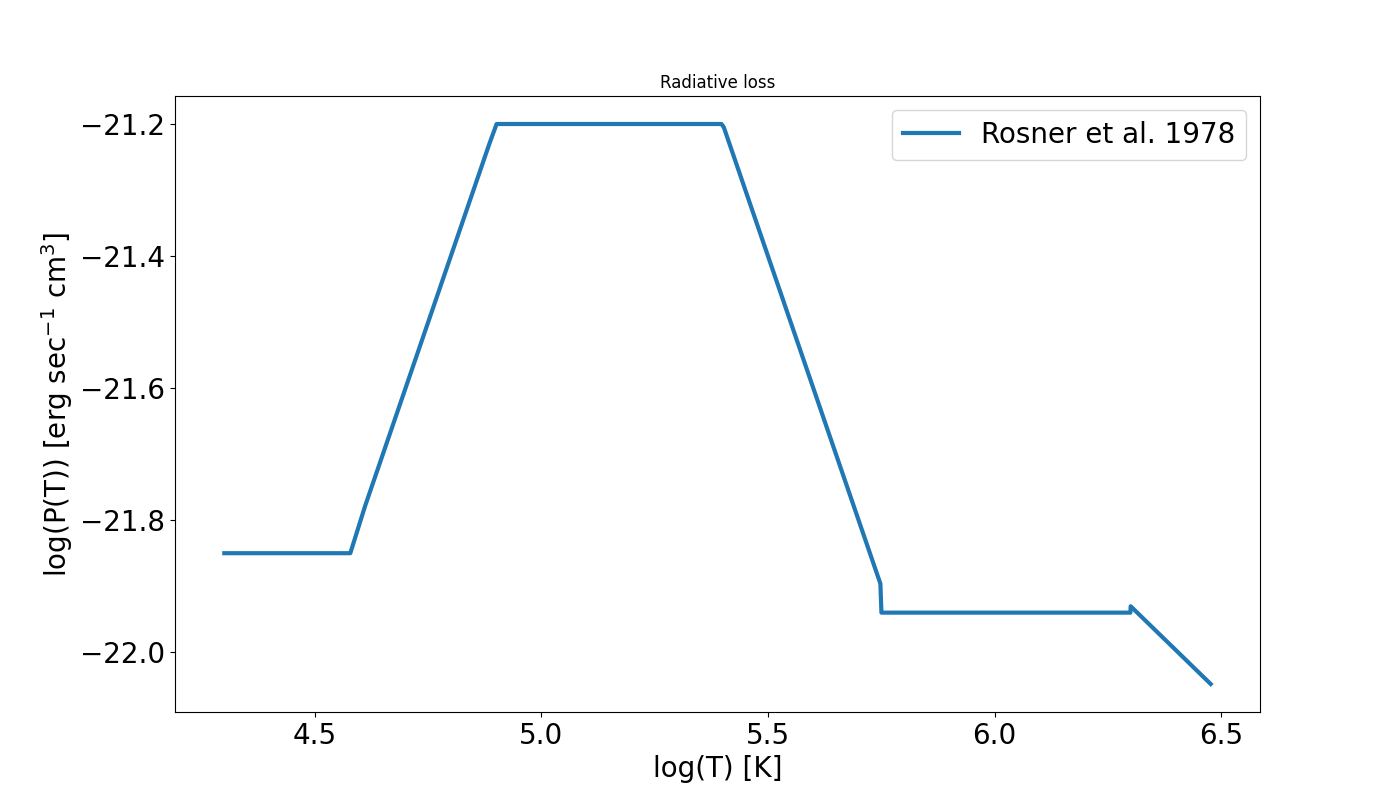}
    \caption{Radiative loss cooling curve defined as in \cite{Rosner1978}. The horizontal axis denotes temperature and the vertical axis shows the radiative loss cooling curve function in cgs. Log-log scale is applied.}
    \label{fig:rad_loss_rosner}
\end{figure}

\begin{figure*}[htb!]
    \centering
    \begin{subfigure}[b]{0.32\textwidth}
         \centering
         \includegraphics[width=\textwidth]{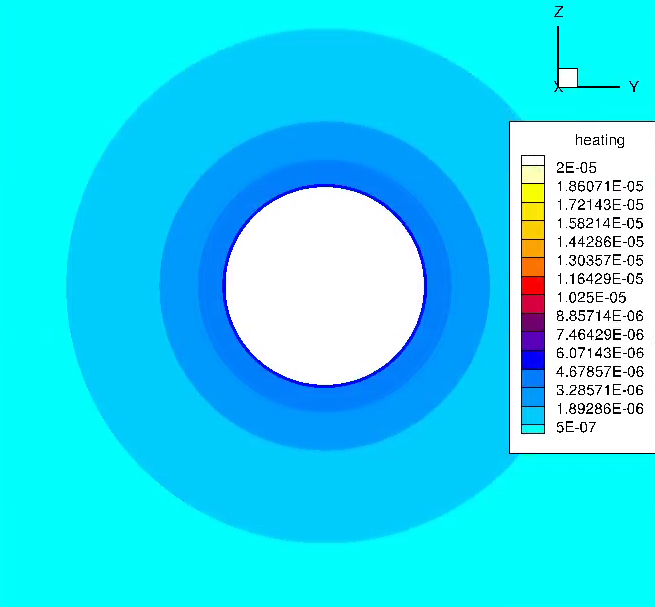}
         
     \end{subfigure}
     \hfill
     \begin{subfigure}[b]{0.32\textwidth}
         \centering
         \includegraphics[width=\textwidth]{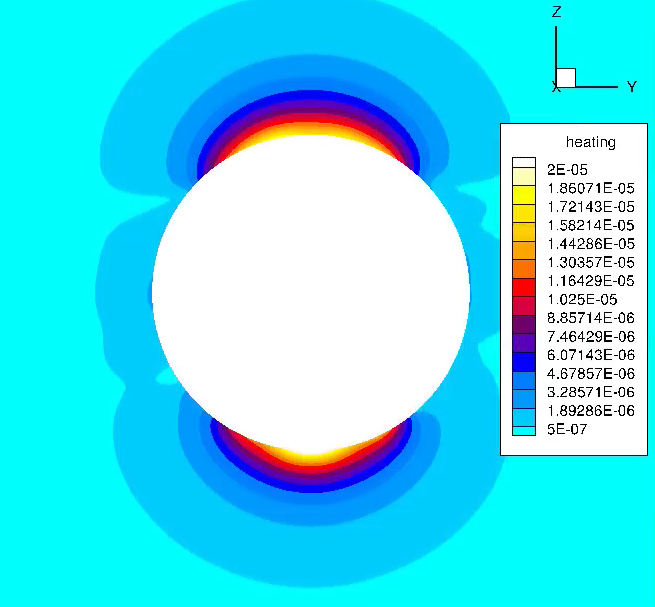}
    
     \end{subfigure}
     \hfill
     \begin{subfigure}[b]{0.32\textwidth}
         \centering
         \includegraphics[width=\textwidth]{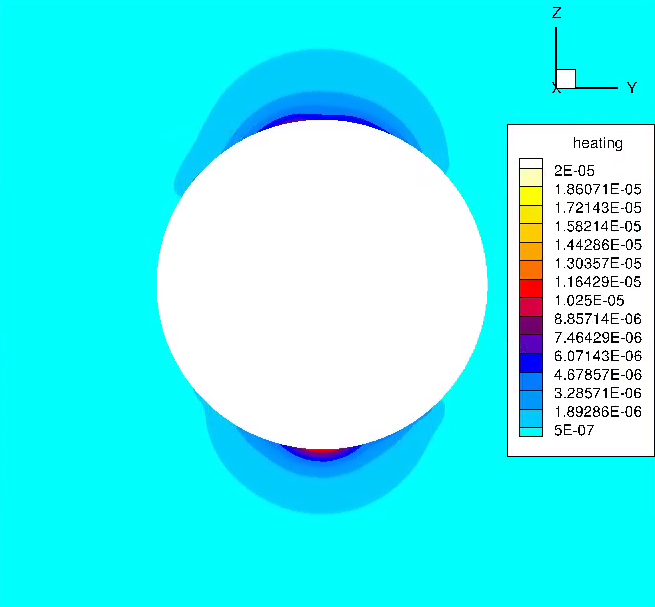}

     \end{subfigure}
     \hfill
    \caption{Obtained heating profiles in the meridional plane for the full MHD COCONUT simulations for each heating description given in Eqs.~\ref{exp_heating},~\ref{magnetic_damping}, and~\ref{heating_lionello}, from left to right, respectively, in [J m$^{-3}$ s$^{-1}$].}
    \label{fig:cr2219_heating_profiles}
\end{figure*}

\begin{figure*}[htb!]
    \centering
    \begin{subfigure}[b]{0.32\textwidth}
         \centering
         \includegraphics[width=\textwidth]{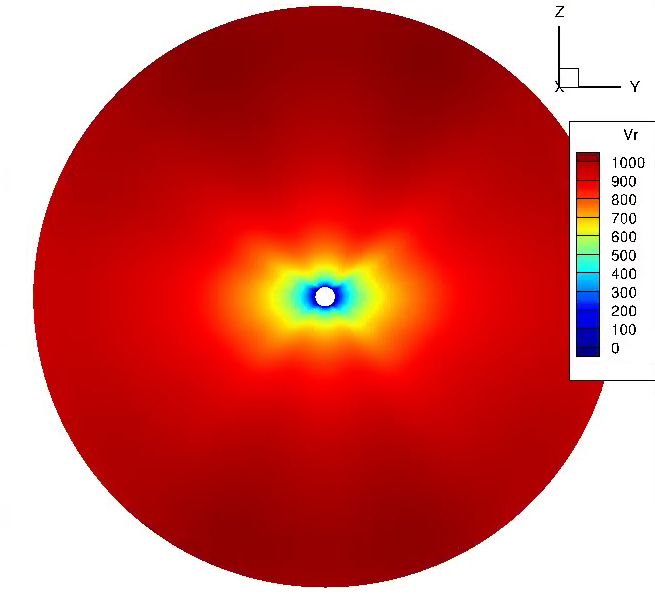}
        
     \end{subfigure}
     \hfill
     \begin{subfigure}[b]{0.32\textwidth}
         \centering
         \includegraphics[width=\textwidth]{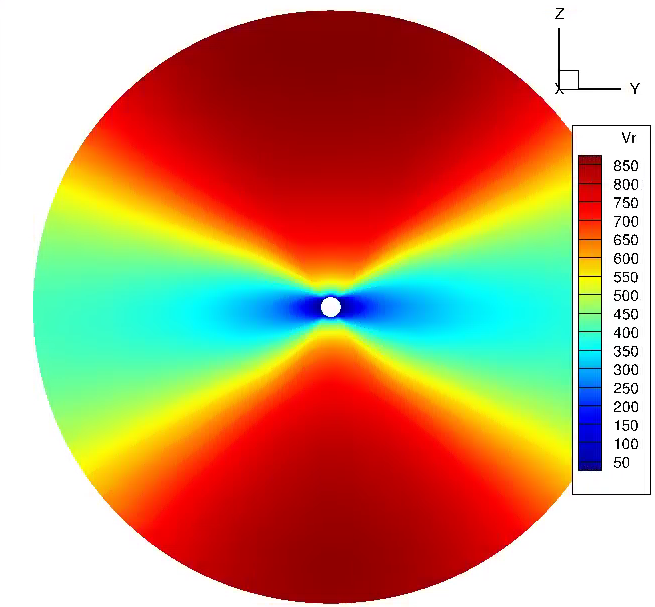}
         
     \end{subfigure}
     \hfill
     \begin{subfigure}[b]{0.32\textwidth}
         \centering
         \includegraphics[width=\textwidth]{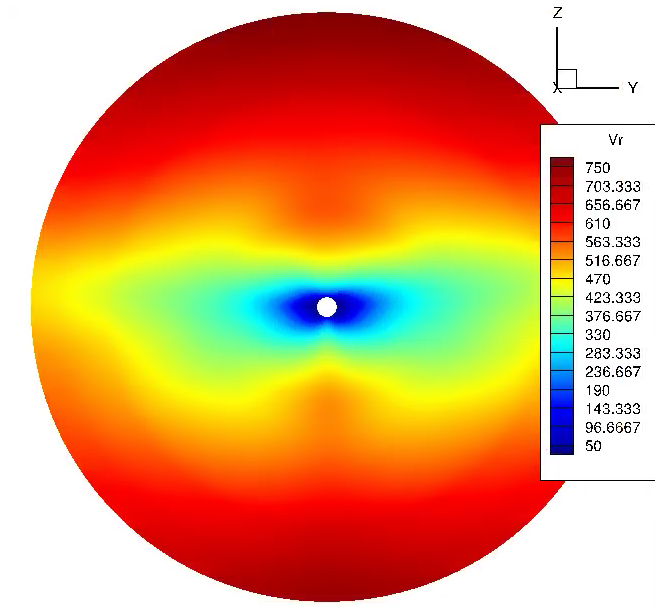}
         
     \end{subfigure}
     \hfill    
    \caption{Radial velocity values in [km s$^{-1}$] in the meridional plane for the full MHD COCONUT simulations with the indicated heating functions given in Eqs.~\ref{exp_heating},~\ref{magnetic_damping}, and \ref{heating_lionello}, from left to right, respectively.} 
    \label{fig:cr2219_radial_velocity_values}
\end{figure*}

\begin{figure*}[hbt!]
     \centering
     \begin{subfigure}[b]{0.32\textwidth}
         \centering
         \includegraphics[width=\textwidth]{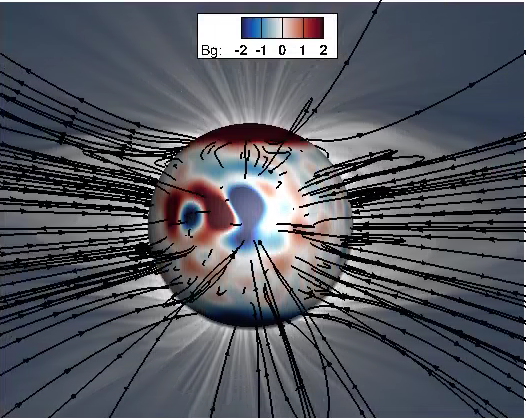}
         
     \end{subfigure}
     \hfill
     \begin{subfigure}[b]{0.32\textwidth}
         \centering
         \includegraphics[width=\textwidth]{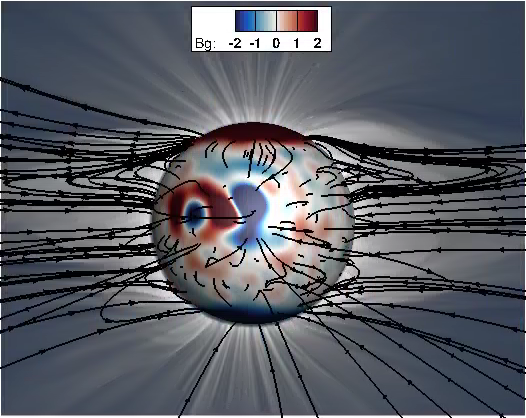}
        
     \end{subfigure}
     \hfill
     \begin{subfigure}[b]{0.32\textwidth}
         \centering
         \includegraphics[width=\textwidth]{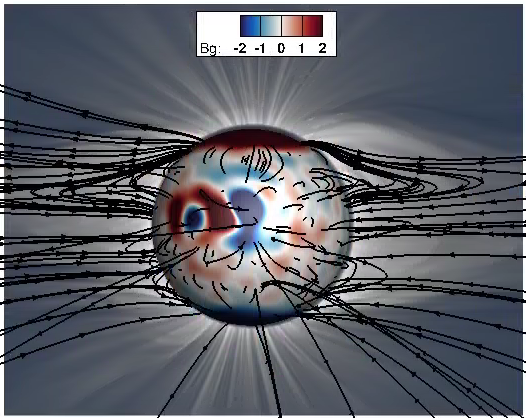}
       
     \end{subfigure}
     \hfill
    \caption{Full MHD COCONUT solutions with the indicated heating profiles given in Eqs.~\ref{exp_heating},~\ref{magnetic_damping}, and \ref{heating_lionello}, from left to right, respectively. The eclipse image is overlaid with the magnetic stream-lines from the COCONUT solution. The radial magnetic field is plotted on the solar surface. }
        \label{fig:cr2219_fieldlines}
\end{figure*}
Thermal conduction was implemented the same way as in \cite{Mikic1999} following \cite{Hollweg1978}. We defined two regimes where plasma is collisional and collisionless. We define thermal conduction as follows:
\begin{equation}
    \vec{q_1} = -\kappa_{||} \vec{\hat{b}} \vec{\hat{b}} \cdot \nabla T
,\end{equation}
where $\vec{q_1}$ is the standard Spitzer-H$\ddot{\text{a}}$rm thermal conduction flux in the collisional regime below 10 R$_\odot$ with $\kappa_{||} = 9\times10^{-7} T^\frac{5}{2}$ in cgs units, as follows:\ 
\begin{equation}
    \vec{q_2} = \alpha n_e k T \vec{v}
,\end{equation}
where $\vec{q_2}$ is the thermal conduction flux in the collisionless regime beyond 10 R$_\odot$, $\alpha$ is a constant given in \cite{Hollweg1978} and $k$ is the Boltzmann constant. Radiative loss is defined in the optically thin limit according to \cite{Rosner1978} by:
\begin{equation}
    Q_{rad} = - n_en_p P(T),
\end{equation}
where $n_e$ and $n_p$  correspond to electron and proton number densities, and it is assumed that $n_e = n_p$ for the hydrogen plasma, $P(T)$ is defined in \cite{Rosner1978} and is a cooling curve depending on the temperature. The defined radiative loss function profile is given in Fig.~\ref{fig:rad_loss_rosner} in log-log scale.

\subsection{Empirical coronal heating} \label{coronal_heating}
The mechanism that heats up the corona is still unresolved in solar physics. It is assumed that magnetic energy is transformed into thermal energy, but the exact mechanism still has to be identified. The coronal heating has been approximated with physics-based models but also with phenomenological approaches (\cite{Schrijver1985}, \cite{Fisher1998}). In our model, we use the three most common approximations found in the literature. As a first approximation, we used an exponential envelope function similar to that of \cite{Lionello2009} and \cite{Downs2010},
\begin{equation} \label{exp_heating}
    Q_{H} = H_0 e^{-\frac{r-R_s}{\lambda}},
\end{equation}
where $R_s$ is the solar radius, $H_0$ is the local heating rate at $r=R_s$ and $\lambda$ is the scale height. We use $H_0 = 4.9 \cdot 10^{-5}\;$erg cm$^{-3}$ s$^{-1}$ and $\lambda = 0.7\;$R$_\odot$.
\\
\cite{Pevtsov2003} established that there is a linear dependence of the magnetic field strength and X-ray radiance. Therefore, a function approximating this law is tested in COCONUT with a slight modification. \cite{Downs2010} also suggested using the radial damping of the heating term when using the coronal heating model based on the magnetic field configuration. Thus, the second model considered here is expressed as:
\begin{equation} \label{magnetic_damping}
    Q_{H} = H_0 \cdot |\mathbf{B}| \cdot e^{-\frac{r-R_s}{\lambda}},
\end{equation}
where $H_0 = 4 \cdot 10^{-5}$ erg cm$^{-3}$ s$^{-1}$ G$^{-1}$ and $\lambda=0.7R_\odot$.
 
Finally, we also considered the most complex heating function approximation that considers exponential heating, which is the contribution describing the quiet Sun and active region heating. The last approximated heating function is taken from \cite{Lionello2009}. The function depends on the magnetic field, approximating the heating for the quiet Sun and the active regions: 
\begin{equation} \label{heating_lionello}
    Q_H = Q_H^{exp} + Q_H^{QS}+Q_H^{AR},
\end{equation}
\begin{equation}
    Q_H^{exp} = H_0 e^{\frac{-(r-R_\odot)}{\lambda_0}},
\end{equation}
where $H_0 = 4.9128 * 10^{-7}$ erg cm$^{-3}$ s$^{-1}$ and $\lambda_0 = 0.7 R_\odot$.
\begin{equation}
    Q_h^{QS} = H_0^{QS}f(r)\frac{B_t^2}{B(|B_r|+B_r^c)},
\end{equation}
where $B_t = \sqrt{B_\theta^2 + B_\phi^2}$ is the tangential magnetic field, and
\begin{equation}
    Q_H^{AR} = H_0^{AR} g(B) \Big( \frac{B}{B_0} \Big)^{1.2}.
\end{equation}
In the performed simulations, the following constants are fixed according to \cite{Lionello2009}, $H_0^{QS} = 1.18 * 10^{-5}$ erg cm$^{-3}$ s$^{-1}$, $B_r^c = 0.55 G$, $H_0^{AR} = 1.87 * 10^{-5}$ erg cm$^{-3}$ s$^{-1}$, $B_0 = 1 G $. The functions f(r) and g(B) are defined as follows:
\begin{equation}
    f(r) = \frac{1}{2} \Bigg(1+\tanh \frac{1.7-r/R_\odot}{0.1} \Bigg) \exp \Bigg(-\frac{r/R_\odot - 1}{0.2} \Bigg),
\end{equation}
\begin{equation}
    g(B) = \frac{1}{2} \Bigg(1+\tanh \frac{B-18.1}{3.97} \Bigg)
,\end{equation}

\begin{figure*}[hbt!]
    \centering
    \includegraphics[width=0.32\textwidth]{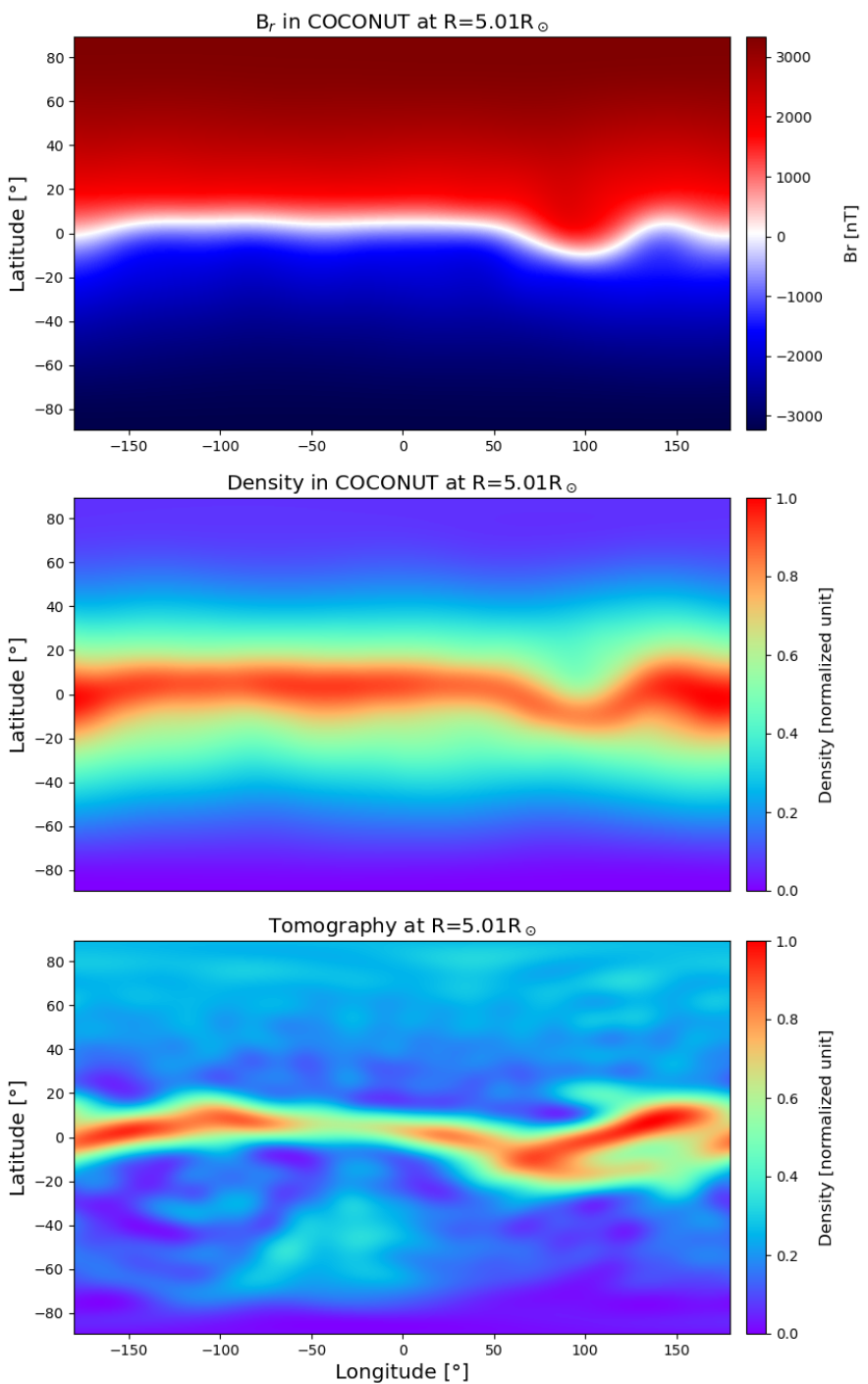}
    \includegraphics[width=0.32\textwidth]{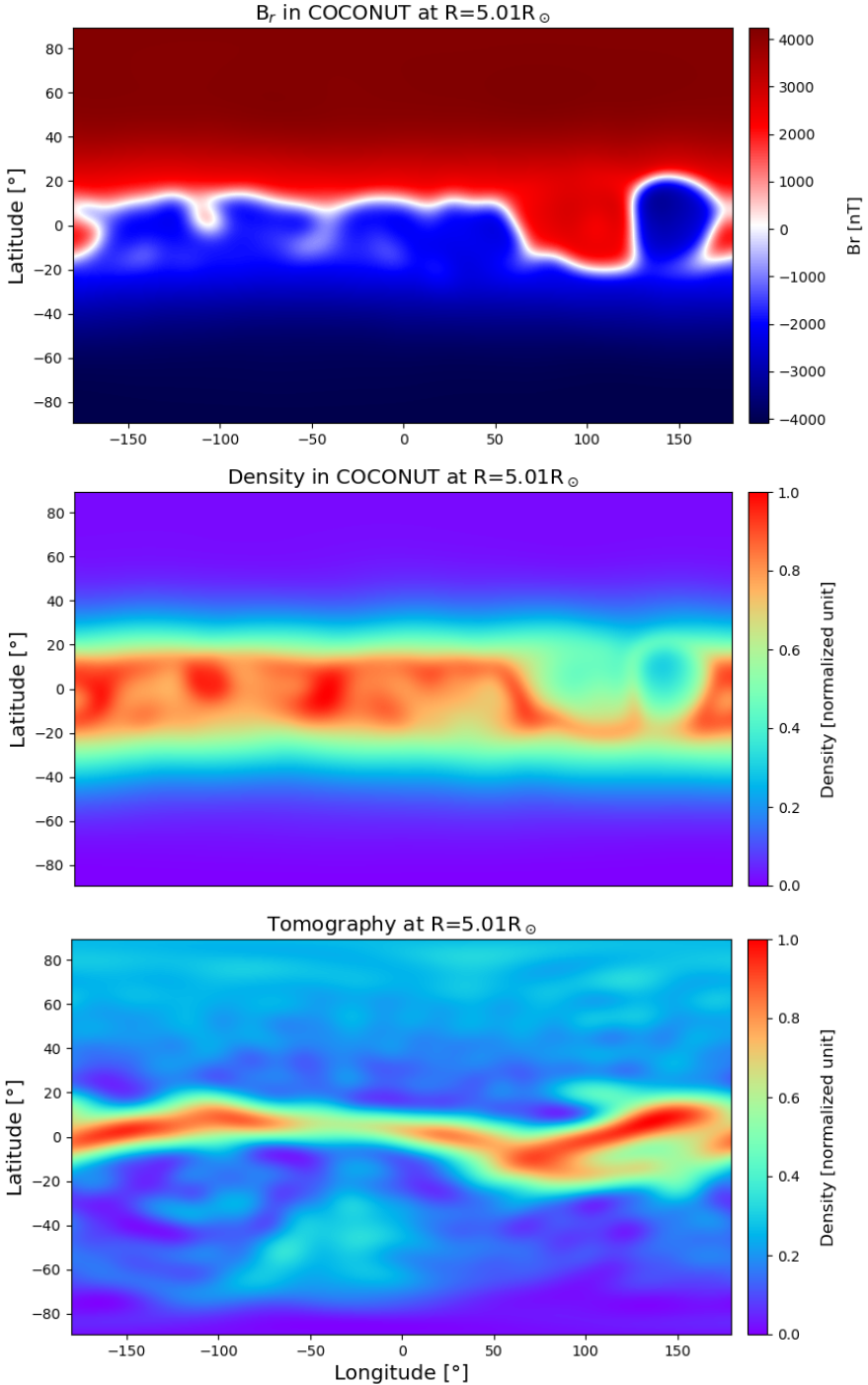}
    \includegraphics[width=0.32\textwidth]{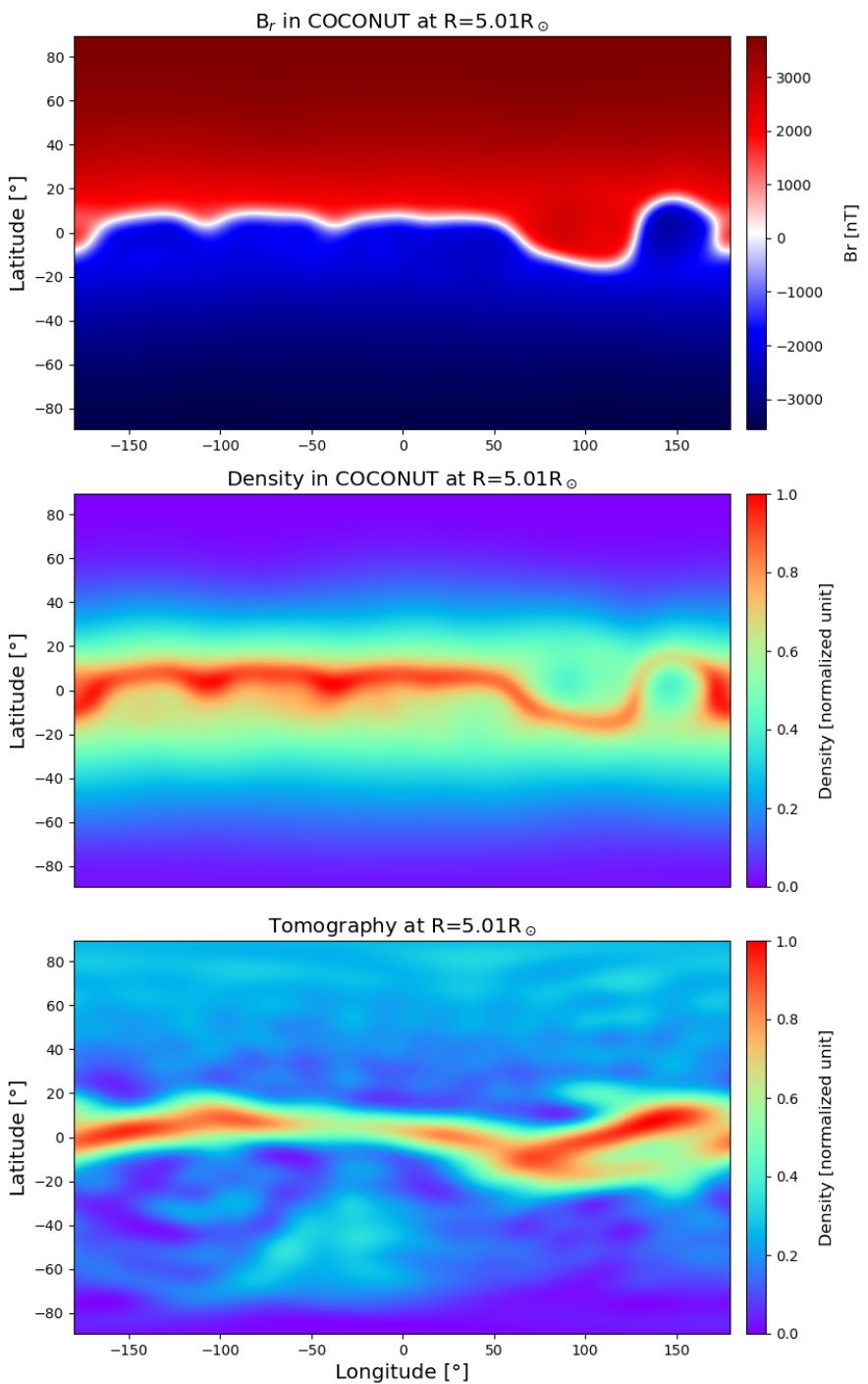}
    \caption{ Longitudes in degrees shown on the horizontal axis. The vertical axis gives the co-latitudes in degrees. Here, $B_r$ is given in [nT], the density from COCONUT in normalised units, and tomography in normalised units, from top to bottom, respectively. All quantities are plotted at $5\;$R$_\odot$. The left panel shows results for the COCONUT simulation with the heating function from Eq.~\ref{exp_heating}, the middle panel shows the results with the heating from Eq.~\ref{magnetic_damping}, and the right includes the heating function from Eq.~\ref{heating_lionello}.}
    \label{fig:tomography_Qh4_Qhlio}
\end{figure*}

The heating profile in Eq.~\ref{heating_lionello} takes into account different heating contributions, but with its initial definition in \cite{Lionello2009}, it is specifically tailored for the test case considered in the original paper; therefore, is has required further modification to adjust to the different solar wind configuration. For our case, we changed the cut-off strength for the magnetic field to 4 Gauss, as the magnetic field is weaker. Thus, in COCONUT, we use the following formula for the active region heating:
\begin{equation}
    g(B) = \frac{1}{2} \Bigg(1+\tanh \frac{B-4}{4} \Bigg).
\end{equation}
Figure~\ref{fig:cr2219_heating_profiles} shows the obtained heating profiles in the meridional plane in the COCONUT simulations for each indicated heating function. The values are given in [J m$^{-3}$ s$^{-1}$] units. The left figure shows the exponential heating function given in Eq.~\ref{exp_heating}, and, as expected, the obtained heating profile is uniform and decreases radially outward. The total power injected by this function is $7.13 \cdot 10^{19} \;$J s$^{-1}$. The heating profile introduced by Eq.~\ref{magnetic_damping} is given in the middle panel, where the profile is no longer uniform, as the magnetic field contribution is considered. The total power injected in the corona with this function is $2.88\cdot 10^{18} \;$J s$^{-1}$. The heating profile introduced in Eq.~\ref{heating_lionello} is given in the last panel of Fig.~\ref{fig:cr2219_heating_profiles}. The different heating contributions are separated here, and the heating coming from the strong magnetic field regions is dominant in this case. The total power injected in the corona with this function is $1.44\cdot 10^{21} \;$J s$^{-1}$, which is the highest among the implemented heating profiles. The total power injected by the heating profiles is reported to be of the order of $10^{19} - 10^{21} \;$J s$^{-1}$ in the literature \citep{Lionello2009, Downs2010, reville2020, Parenti2022}.

Figure~\ref{fig:cr2219_radial_velocity_values} shows the obtained heating functions in the meridional plane in the COCONUT simulations for each indicated heating function. The velocity profile near the outer coronal boundary obtained by Eq.~\ref{exp_heating} is uniform but still very fast. The wind profile obtained with Eq.~\ref{magnetic_damping} is bi-modal, as the speed is slower near the equator than at the poles. The solar wind speed profile is also almost bi-modal in the last panel of figure~\ref{fig:cr2219_radial_velocity_values}. However, the slow speed stream is small at 0.1~au in this case. 

Figure~\ref{fig:cr2219_fieldlines} shows the overview of the effect of the different heating profiles in the low corona. In all the figures, the observed eclipse image is plotted in the background and magnetic streamlines are overlaid from COCONUT simulations. The solar surface is coloured with the radial magnetic field values. 

The left figure shows the profiles for the heating function given in Eq.~\ref{exp_heating}. The heating introduced by this equation with the given scale height values is uniform yet strong. The field lines seem to be blown out and radially stretched -- and not collimated towards the equator.  The second panel represents the obtained solar wind with Eq.~\ref{magnetic_damping}. The stream lines are elongated compared to the fieldlines in the polytropic case. Here, the injected heating in the corona was sufficient to speed up the solar wind and obtain the bi-modal structure. The right figure shows the obtained coronal configuration for Eq.~\ref{heating_lionello}. The obtained heating profile also accelerated solar wind sufficiently.
The total power injected by the heating functions in the domain is not overestimated in COCONUT simulations. In the first two cases, it is underestimated. However, we can see that the energy deposition in a specific place plays a crucial role in obtaining a bi-modal wind configuration near the outer heliospheric boundary. For example, the total power injected by Eq.~\ref{magnetic_damping} is less than in the case of Eq.~\ref{heating_lionello}. Nevertheless, the bi-modal structure of the solar wind achieved by the former is clearer than by the latter heating profile. We also notice that the deposited heating strongly affects the loops' shape. A similar effect to the stretched magnetic loops was also obtained in the Wind Predict simulation in \citep{Parenti2022}, where the  authors observed the stretching of the helmet streamers depending on the location and strength of the deposited thermal energy, together with the number density configuration at the base of the corona.

\begin{figure}[hbt!]
    \centering
    \includegraphics[width=0.43\textwidth]{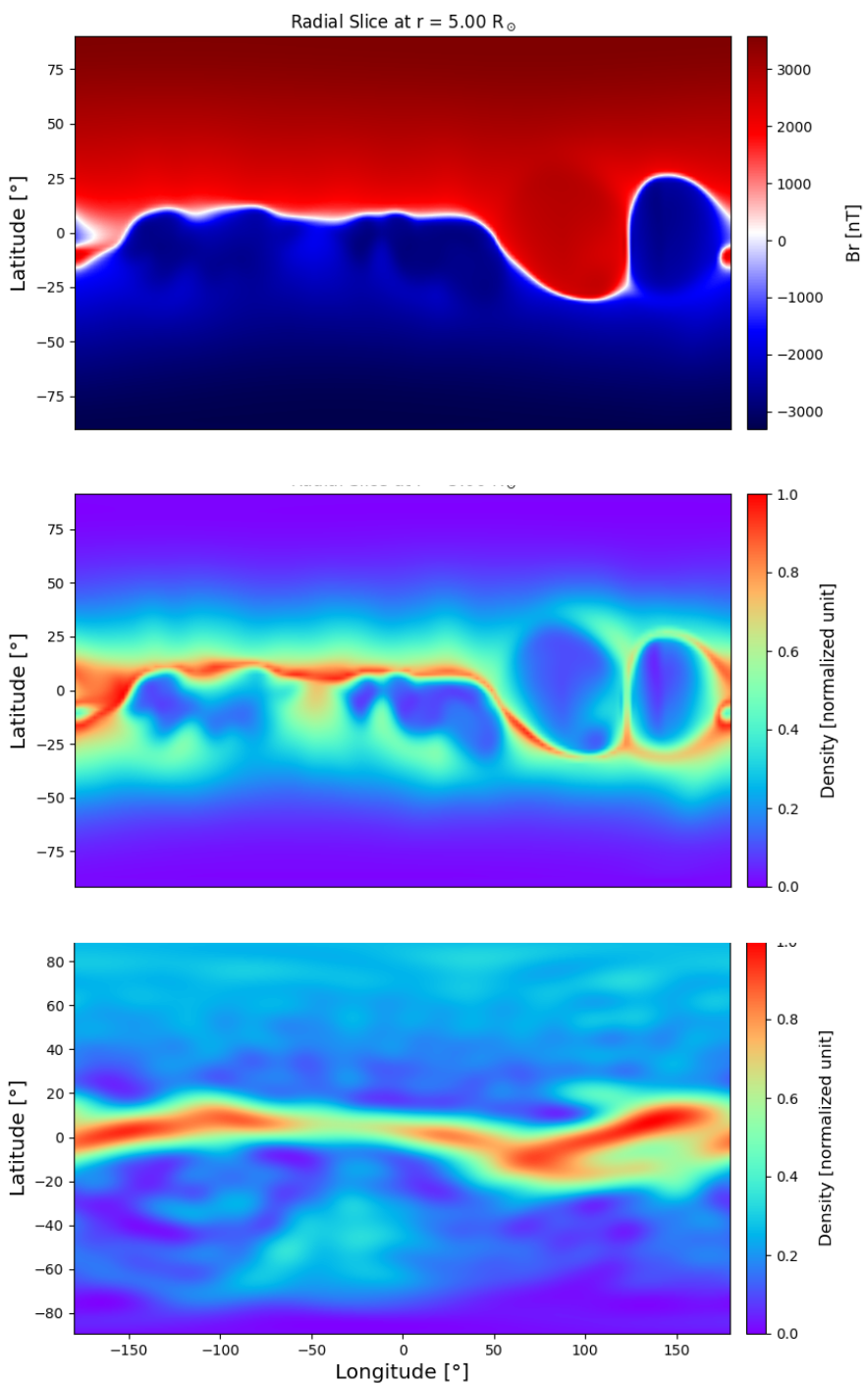}
    \caption{ Longitudes in degrees are shown on the horizontal axis. Vertical axis gives the co-latitudes in degrees. The panels represent Br in [nT], the density from MAS simulations in normalised units, and tomography in normalised units from top to bottom, respectively. All quantities are plotted at $5\;$R$_\odot$. The MAS simulation was obtained with thermodynamic model 1.}
    \label{fig:PSI_thermo1}
\end{figure}
Figure~\ref{fig:tomography_Qh4_Qhlio} shows the radial magnetic field, density and tomography data at 5 R$_\odot$ from the COCONUT simulation with the full MHD model based on heating functions given in Eq.~\ref{exp_heating} in the left panel. The middle panel shows the results with the heating from Eq.~\ref{magnetic_damping} and the right one includes the heating function from Eq.~\ref{heating_lionello}. The structures of the density profiles in all figures show more features compared to the polytropic case (Fig.~\ref{fig:tomography_polytropic}), which again is linked to the current sheet. The density is more enhanced near the equatorial plane in COCONUT simulations than in the tomography data, partially due to the resolution limitation and the prescribed density at the base of the corona. 
Nevertheless, the distinctive features present in the simulation results (middle row) between [$50^\circ - 180^\circ$] degrees in longitude coincide with the observation tomography data (bottom row), as the enhanced density surrounds the lower density region. The distinct contributions from the heating function in Eq.~\ref{heating_lionello} modelled more realistic density profiles in the corona, based on the comparison of the number density profiles between [$50^\circ - 180^\circ$] degrees in longitude. The shape of the high values in the density profiles; namely, the high-density profiles that appear at $~10^\circ$ latitude and $50^\circ$ longitude and shift towards $+20^\circ$ latitude as the longitude values increase towards $180^\circ$ degrees are more similar to the tomography observation data plotted in the bottom panel. This signature is absent in the result obtained with the heating profile given in Eq.~\ref{magnetic_damping}. The number density modelled by the heating profile in Eq.~\ref{exp_heating} demonstrates a similar profile to the tomography data. The enhanced density is more localised near the equator. However, the prominent feature of the lower-density region near the equator surrounded by the higher-density region is missing. Figure~\ref{fig:PSI_thermo1} shows the radial magnetic field and number density at 5 R$_\odot$ from MAS simulation with the thermodynamic model 1 available on their website. This model includes the transition region and the wave turbulence model, in contrast to the COCONUT simulation. We can see that the features are present in the same place compared to Fig.~\ref{fig:tomography_Qh4_Qhlio}, and the density enhancement is also more prominent than the tomography data. The result of the complex heating function implemented in COCONUT yielded a more similar number density profile to the MAS simulation results, which could be because the additional heating in MAS is introduced with the same formula from \cite{Lionello2009}.
 
Table~\ref{table:run_times} summarises the computational resources spent on the simulations. All COCONUT simulations were performed on four nodes of the Genius cluster of Vlaams Supercomputing Center. During the iterative process, the run times were taken when the residuals for the density components reach 10$^{-3}$. The first row represents the polytropic simulation, which takes $\sim 0.98\;$h. The second-to-last rows give the results for the full MHD COCONUT simulations with different heating profile approximations. The heating profile introduced by Eq.~\ref{exp_heating} required $0.7\;$h to converge to the steady-state solution. The heating profile given by Eq.~\ref{magnetic_damping} that resulted in the successful bi-modal solar wind took 1.2 h, which is $\sim 0.2$ h longer than the polytropic simulation. The heating profile approximated by Eq.~\ref{heating_lionello} took 1.94 hours to converge, which is only $\sim 1$ h slower than the polytropic simulation. As a result, obtaining the bi-modal solar wind by activating the full MHD source terms does not require a significantly increased computational time, making it suitable for space weather forecasting.

\begin{table}[htb!]
  \caption{Run times (wall-clock time) required for the COCONUT simulations to reach convergence -3 in density. All the simulations were performed on four  nodes with two Xeon Gold 6240 CPUs@2.6 GHz (Cascadelake), 18 cores each, on the Genius cluster at KU Leuven.}
  \centering
   \begin{tabular}{c c  }
  \hline\hline
   Simulation & Time \\[4pt]
   \hline
Polytropic & 0.98 h \\[4pt] 
\hline
Eq.~\ref{exp_heating} & 0.7 h \\[4pt] 
\hline
Eq.~\ref{magnetic_damping} & 1.2 h \\[4pt] 
\hline
Eq.~\ref{heating_lionello} & 1.94 h \\[4pt] 
\hline
 \end{tabular}
  \label{table:run_times}

\end{table}

\section{Coupling to Icarus} \label{coupling_icarus}
\begin{figure*}[htb!]
    \centering
    \includegraphics[width=0.32\textwidth]{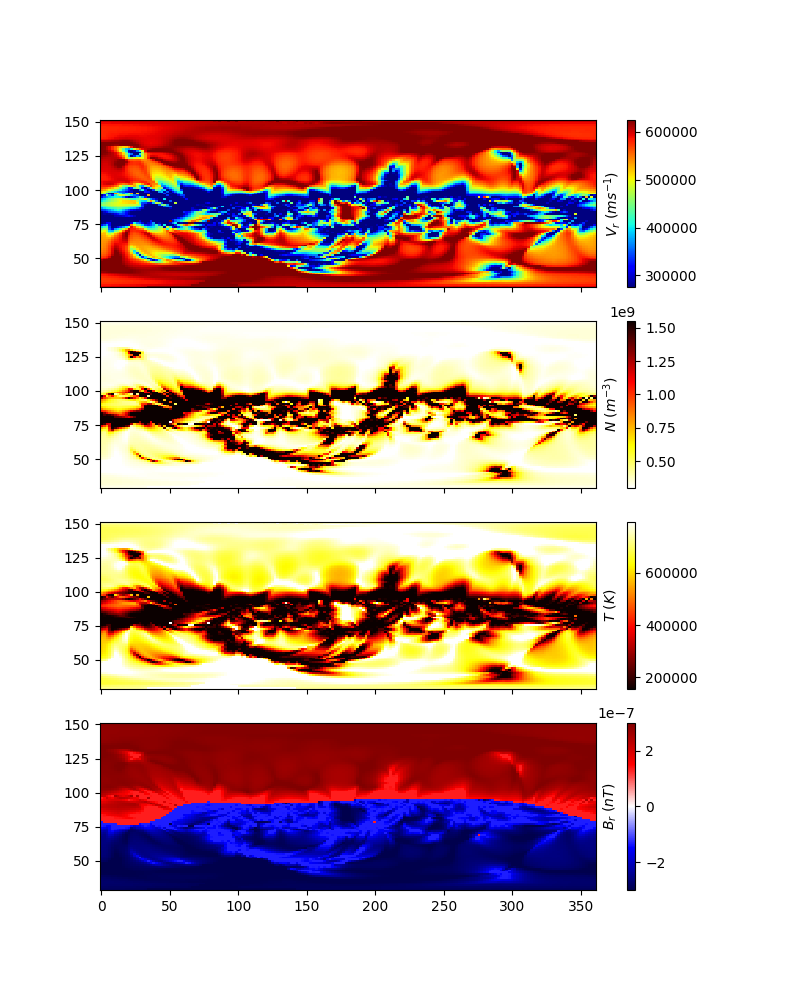}
    \includegraphics[width=0.32\textwidth]{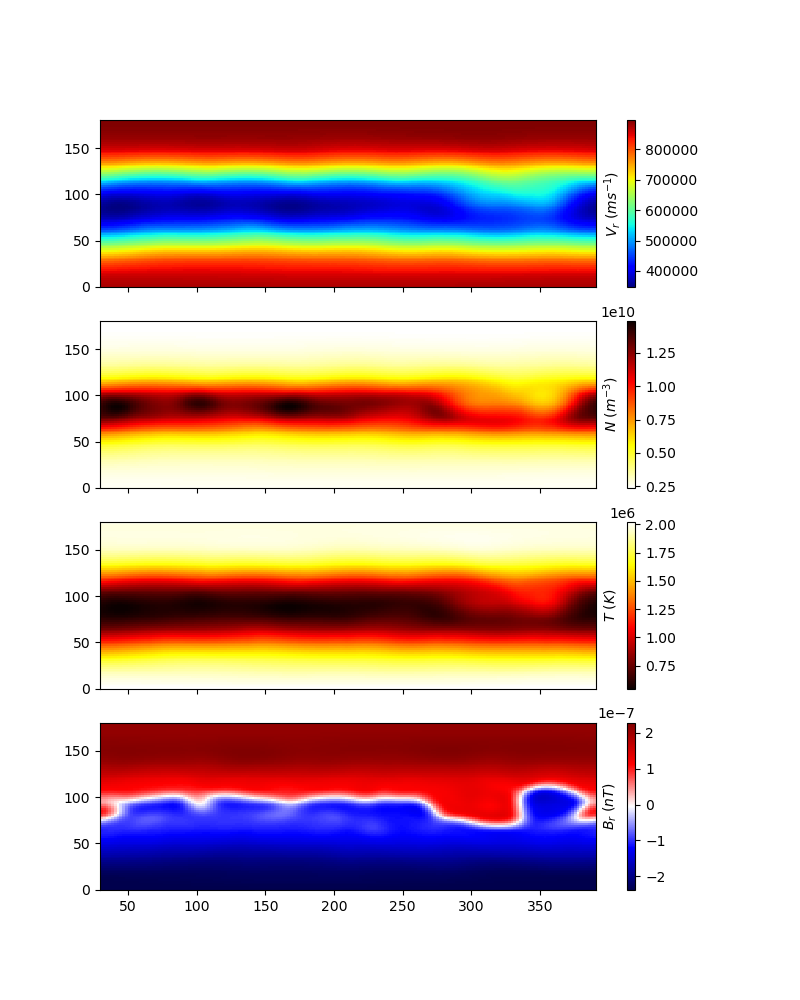}
    \includegraphics[width=0.32\textwidth]{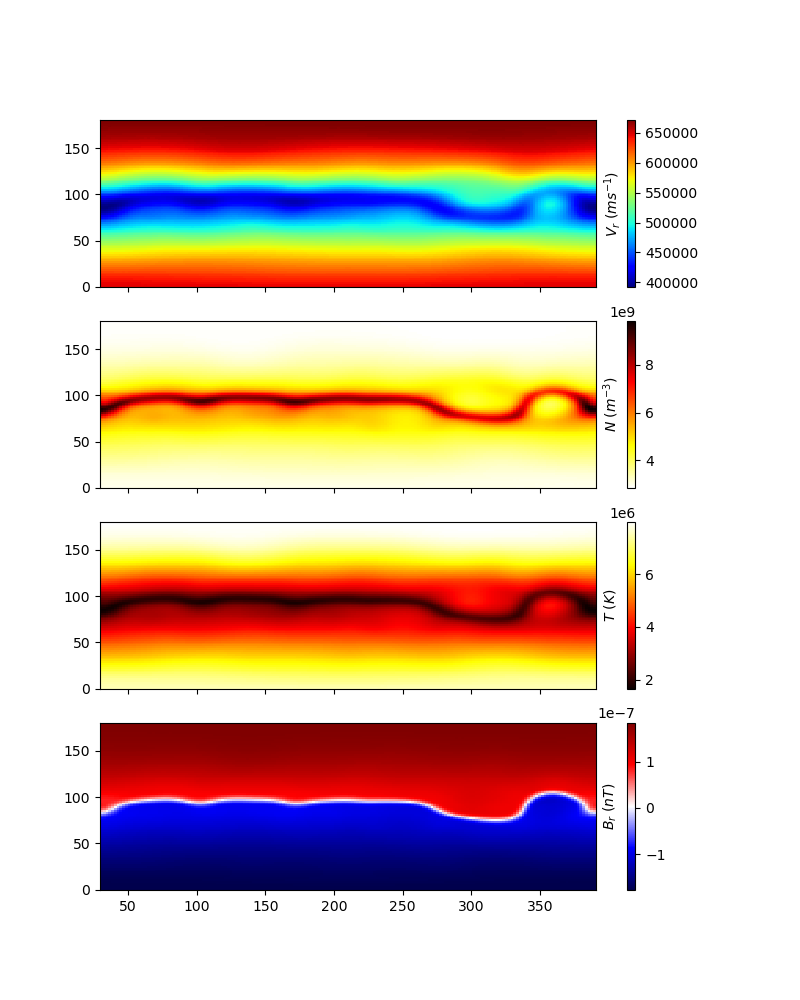}
    \caption{Input boundary files generated from the WSA coronal model (left panel), COCONUT simulation with heating Eq.~\ref{magnetic_damping} (middle panel), and COCONUT simulation with heating Eq.~\ref{heating_lionello} (right panel). The horizontal axis shows longitudes, and the vertical axis shows latitudes. The variables are given from top to bottom as follows: radial velocity in m s$^{-1}$, number density in m$^{-3}$, temperature in K, and radial magnetic field in T. }
    \label{fig:boundary_files}
\end{figure*}

\begin{figure*}[hbt!]
     \centering
     \includegraphics[width=0.33\textwidth]{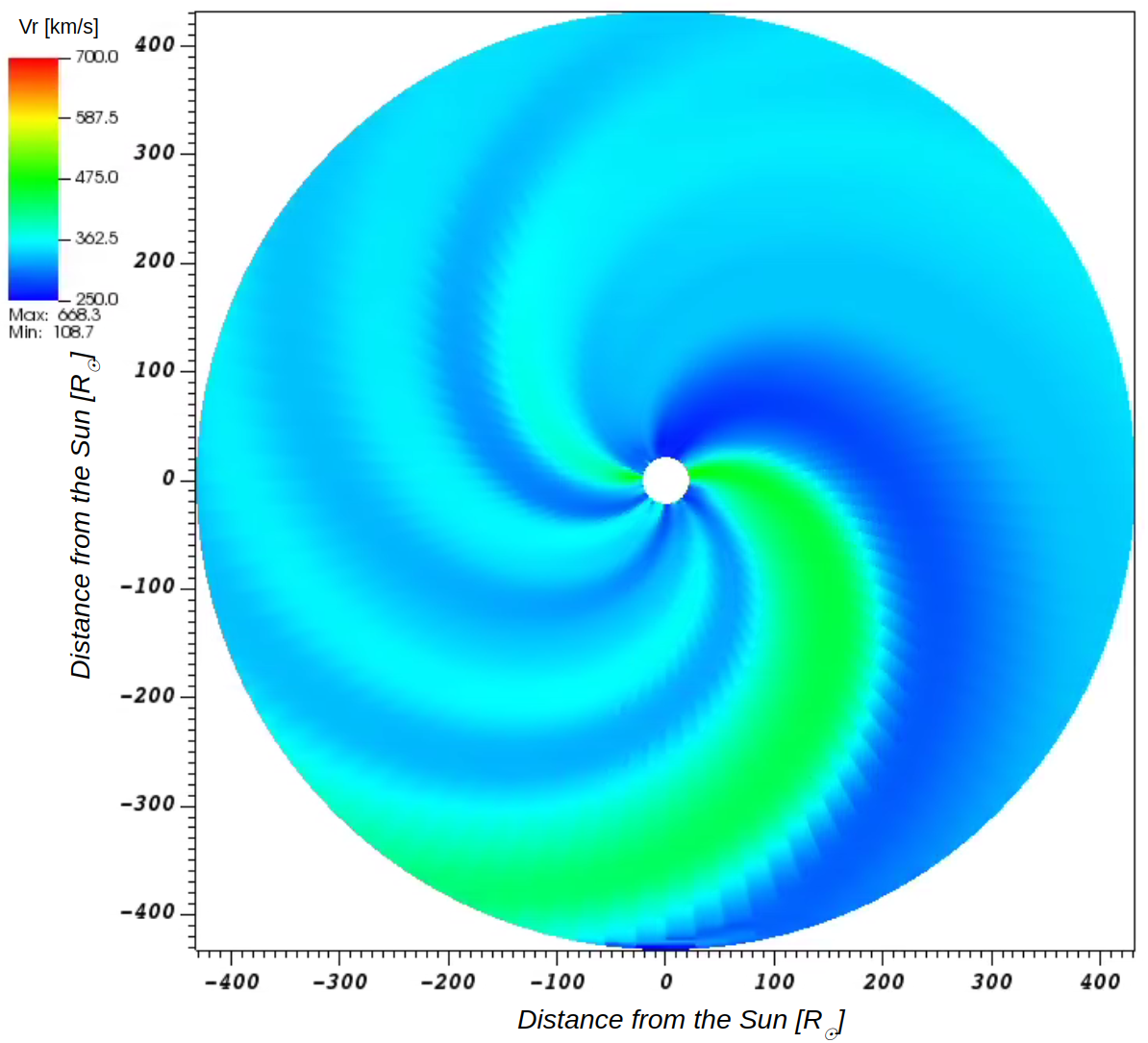}
     \includegraphics[width=0.33\textwidth]{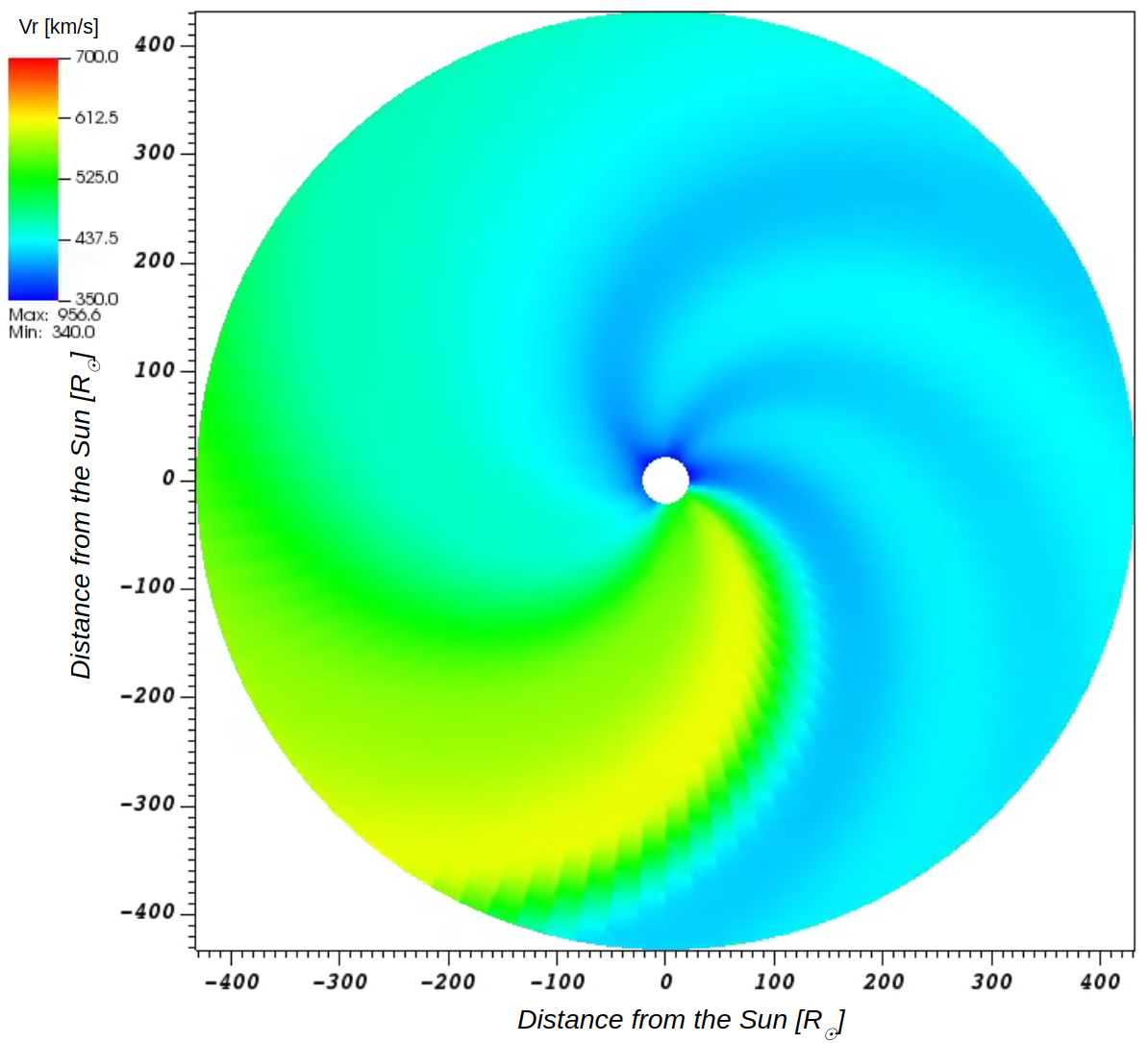}
     \includegraphics[width=0.33\textwidth]{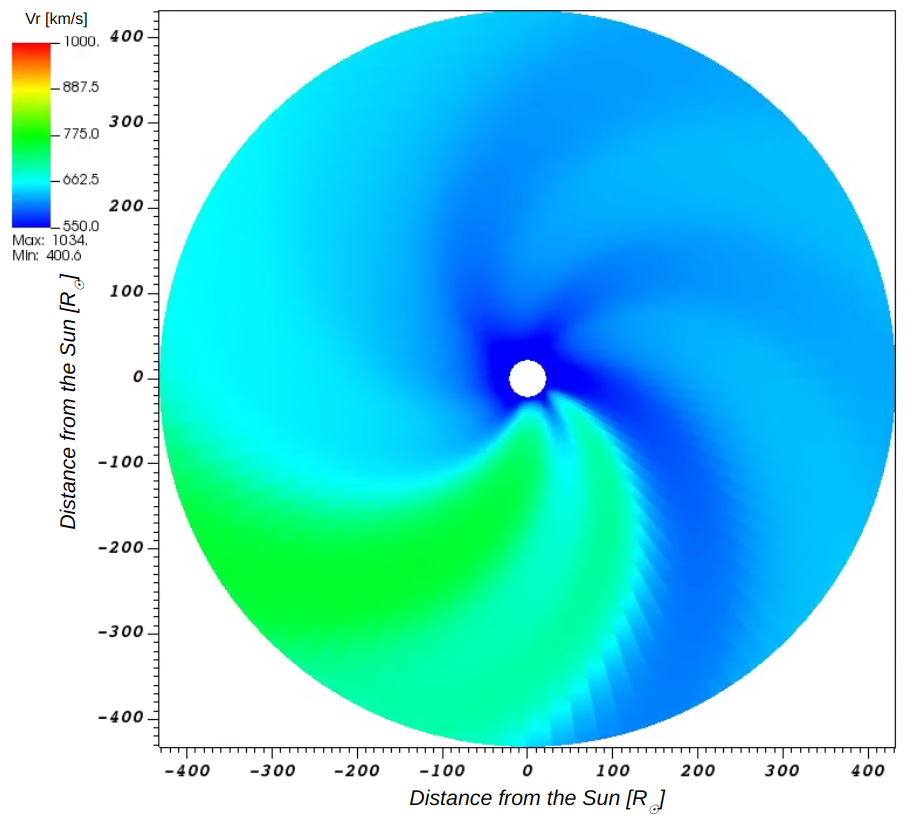}
    \caption{Equatorial plane, showing radial velocity values in Icarus simulations using different coronal boundary files from left to right: from the WSA model, from the COCONUT simulation with the heating function defined in Eqs.~\ref{magnetic_damping} and in~\ref{heating_lionello}. Radial velocity values are given in [km s$^{-1}$]. }
        \label{fig:icarus_equatorial_vr_plots}
\end{figure*}

Icarus is a 3D MHD heliospheric modelling tool (\cite{Verbeke2022}, \cite{Baratashvili2022}) developed in the framework of MPI-AMRVAC \citep{Xia2018}. Icarus was developed as an alternative heliospheric model within the EUHFORIA space weather forecasting chain \citep{Pomoell2018}. It covers the radial distances starting from 0.1~AU to 2~AU, including the orbit of Mars. It uses the finite volume TVDLF solver and a second-order slope limiter. The details of the numerical methodology are given in \cite{Baratashvili2022sungeo}. Icarus is a highly optimised heliospheric tool, as it uses advanced techniques, such as grid stretching and adaptive mesh refinement (AMR),  to obtain results quickly at the desired resolution. Different AMR techniques were discussed in \cite{Verbeke2022} and \cite{Baratashvili2022}, as the user has the freedom to tailor the AMR condition according to the purpose of the run. Icarus uses the output of the current semi-empirical coronal model as inner boundary conditions for driving the solar wind at 0.1~AU. The reference frame is co-rotating with the Sun; thus, after the relaxation period passes (the time that is required for the slow solar wind to traverse the heliosphere from 0.1~AU to the outer boundary at 2~AU), the solar wind is stationary. After this, the CMEs can be injected from the inner heliospheric boundary. Currently, the simple hydrodynamics cone CME model and the magnetised linear force-free spheromak models are supported \citep{Baratashvili2023}. Thus, Icarus can be used to study the propagation of the CMEs, its interaction with the solar wind and the effect on different satellites. Planets and satellites are included in the heliosphere, where plasma and magnetic field characteristic variables are sampled. Currently, there is Mercury, Venus, Mars, Earth, Stereo A, Stereo B, Solar Orbiter, and Parker Solar Probe sampling included. After performing the simulation, the observed data from the satellites can be compared to the sampled data at the satellite locations.
 
The coupling of COCONUT and Icarus was not straightforward, as COCONUT uses an unstructured grid and Icarus uses a structured grid. Data were extracted from the converged solution in COCONUT. It was interpolated, and data corresponding to 0.1~AU was extracted. Then, the data were stored similarly as the semi-empirical WSA coronal model output, used both by the original EUHFORIA heliosphere and Icarus.

Figure~\ref{fig:boundary_files} shows the input boundaries generated from the WSA semi-empirical coronal model of EUHFORIA, the full MHD COCONUT coronal model with heating from Eq.~\ref{magnetic_damping}, and the full MHD COCONUT coronal model with heating from Eq.~\ref{heating_lionello}. The variables plotted from top to bottom are radial velocity, number density, temperature, and radial magnetic field. The horizontal axis shows longitudes, and the vertical axis shows latitudes in degrees. The values between (30$^\circ$, 150$^\circ$) are shown on the vertical axis because the heliospheric models (i.e. the original EUHFORIA heliosphere and Icarus) do not have poles and extend from -60$^\circ$ to 60$^\circ$ in latitudes. In the simulation, the latitudes are transformed to co-latitudes, where the equator corresponds to $90^\circ$ degrees. The results from the WSA model and COCONUT models are quite different; however, the results from two different COCONUT simulations are indeed similar. The density and temperature are strongly overestimated in COCONUT simulations. 

\begin{figure}[htb!]
    \centering
    \includegraphics[width=0.5\textwidth]{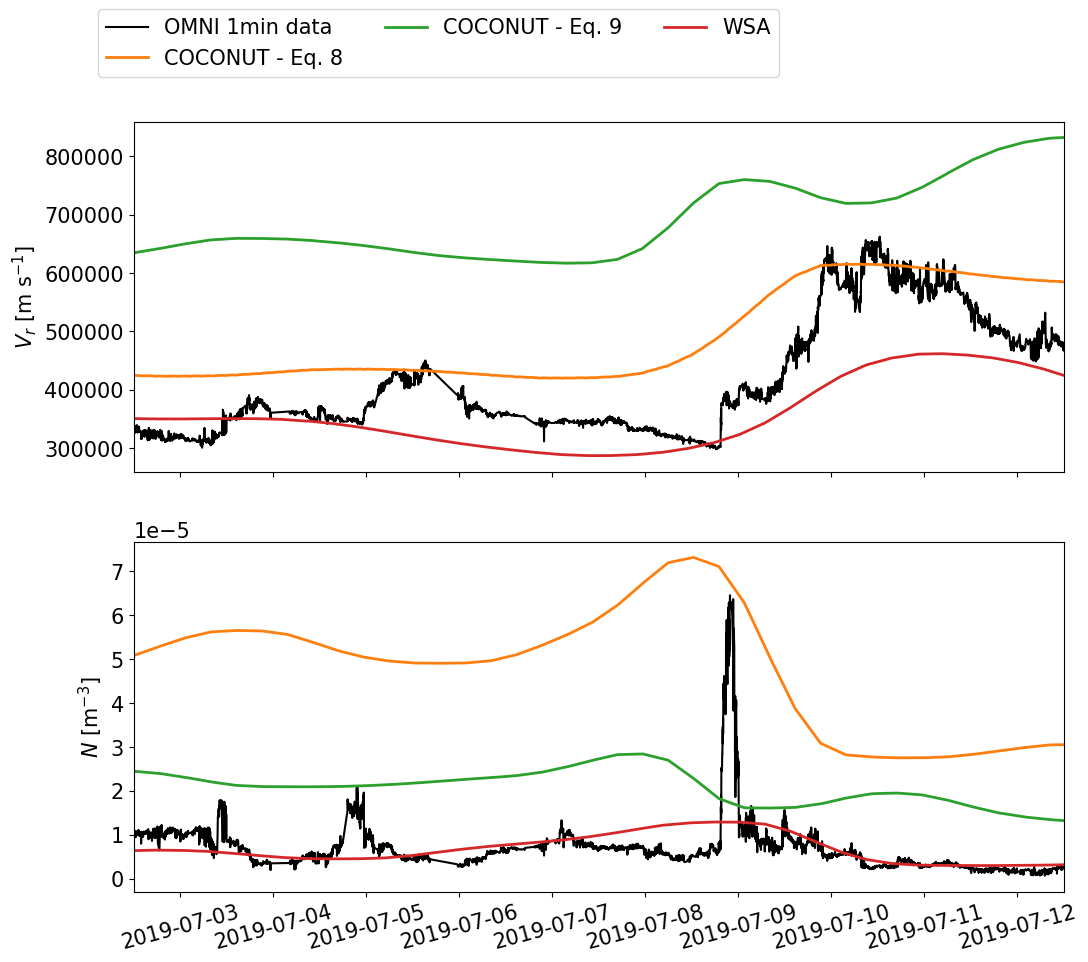}
    \caption{Time series obtained at Earth. The horizontal axis shows longitudes, while the vertical axis shows latitudes. The radial velocity [m s$^{-1}$] and number density [m$^{-3}$] values are plotted. The black curve corresponds to the OMNI 1-min data. Green, red, and orange curves correspond to modelled data from Icarus with the following input boundary files: COCONUT Eq.~\ref{heating_lionello}, WSA model, and COCONUT (Eq.~\ref{magnetic_damping}).}
    \label{fig:speed_n_icarus}
\end{figure}

\begin{figure}[htb!]
    \centering
    \includegraphics[width=0.5\textwidth]{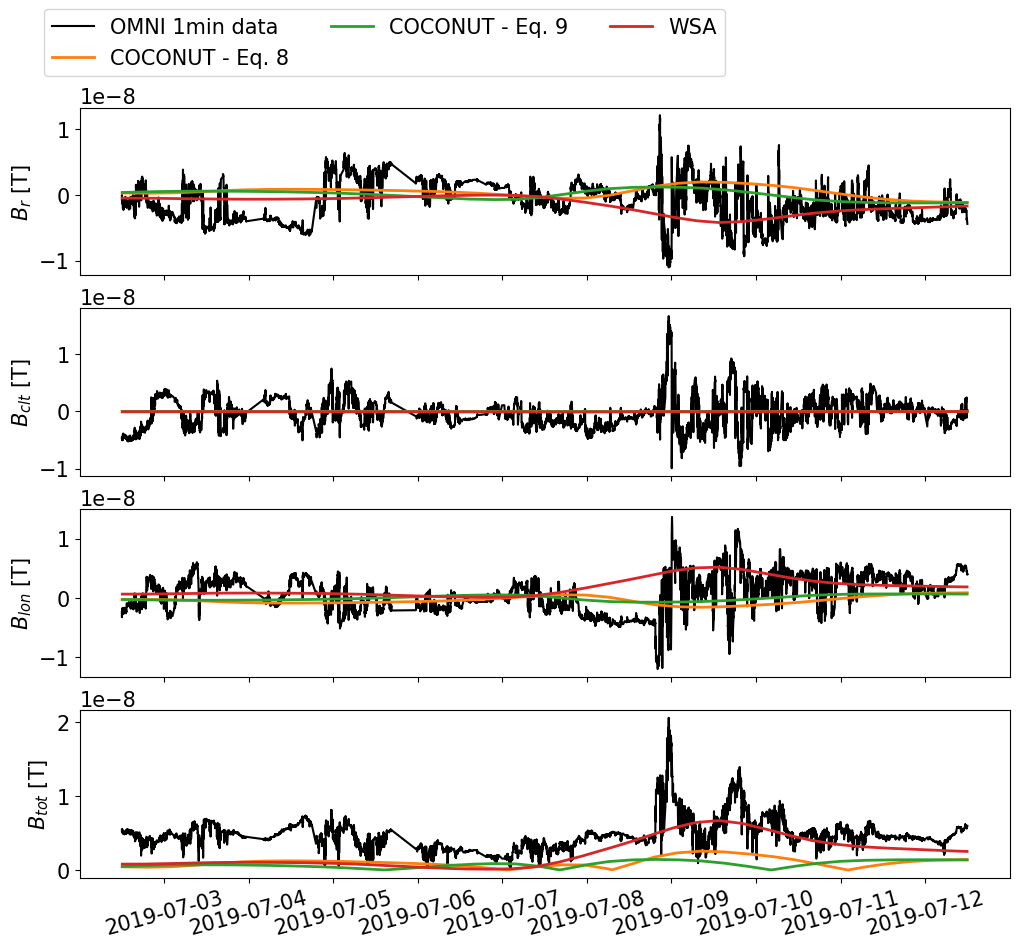}
    \caption{Time series obtained at Earth. The horizontal axis shows longitudes, and the vertical axis shows latitudes. The magnetic field components are plotted in T. The black curve corresponds to the OMNI 1-min data. Green, red, and orange curves correspond to modelled data from Icarus with the following input boundary files: COCONUT Eq.~\ref{heating_lionello}, WSA model, and COCONUT (Eq.~\ref{magnetic_damping}). }
    \label{fig:B_field}
\end{figure}

Figure~\ref{fig:icarus_equatorial_vr_plots} shows the relaxed solutions for the Icarus simulations with the input boundary files demonstrated in Fig.~\ref{fig:boundary_files}. The radial velocity values are plotted in the equatorial plane. The left figure corresponds to the wind modelled by the WSA coronal input file, where the speed ranges between 250–700 km s$^{-1}$. There are multiple higher and lower-speed streams. The middle figure corresponds to the wind simulated by the COCONUT model with the heating function in Eq.~\ref{magnetic_damping}. The speed ranges between 350–700 km s$^{-1}$. The higher speed streams are more diffused into the low-speed streams here. The last figure corresponds to the wind modelled by the COCONUT coronal model with the heating function in Eq.~\ref{heating_lionello}. The speed is considerably higher here, ranging between 550–1000 km s$^{-1}$. The time series obtained at Earth by these three different coronal models are demonstrated in Figs.~\ref{fig:speed_n_icarus} and \ref{fig:B_field}. The former shows the radial velocity and number density values at Earth, whereas the latter shows the magnetic field components. In both figures, the results from the inputs of the WSA coronal model, COCONUT with heating (Eqs.~\ref{magnetic_damping} and \ref{heating_lionello}) are shown in red, orange and green colours, respectively. The black line corresponds to OMNI 1-min data. The WSA input boundary slightly underestimates the speed values compared to observations and models the number density well, compared to the overall profile in the observed data; however, the peak in the observed number density profile is missing from the synthetic data. The COCONUT input boundary with the heating profile presented in Eq.~\ref{magnetic_damping} models the speed range well compared to the observations; however, it overestimates the overall density profile, although the peak density observed in the density can be noticed in the modelled data. The COCONUT input boundary with the heating profile presented in Eq.~\ref{heating_lionello} strongly overestimates the speed values, but it does model the number density better with respect to the observed data than the COCONUT model with the heating profile from Eq.~\ref{magnetic_damping}. The magnetic field values modelled by COCONUT simulations are quite similar. They both are different from the one generated with the WSA input boundary file. In the total magnetic field panel, we can see that COCONUT underestimates the total magnetic field strength compared to the WSA model and the observed data.

\begin{table}[htb!]
  \caption{Run times (wall-clock time)for the heliospheric simulation for 24 days of simulation time in Icarus. All the simulations were performed on four nodes with two Xeon Gold 6240 CPUs@2.6 GHz (Cascadelake), 18 cores each, on the Genius cluster at KU Leuven.}
  \centering
   \begin{tabular}{c c  }
  \hline\hline
   Simulation & Time \\[4pt]
   \hline
WSA input & 5m 56s \\[4pt] 
\hline
Eq.~\ref{magnetic_damping} & 5m 42s \\[4pt] 
\hline
Eq.~\ref{heating_lionello} & 5m 48s \\[4pt] 
\hline
 \end{tabular}
  \label{table:run_times_icarus}
\end{table}

\section{Conclusions and outlook} \label{conclusions}
Our novel full MHD modelling chain was established from the Sun to Earth with COCONUT coronal and Icarus heliospheric models. To use COCONUT for heliospheric modelling, the coronal model was upgraded from the ideal polytropic MHD model to a full MHD model by introducing source and sink terms in the energy equation to account for thermal conduction, radiative cooling and coronal heating. The adiabatic index $\gamma = 5/3$ was used instead of $\gamma = 1.05$. In the first attempt, the coronal heating was approximated by various functions, depending on the radial distance and the magnetic field strength in the corona. The implemented full MHD coronal model was compared to the coronal model of MAS. This was done by comparing the bi-modal structure of the wind in the meridional plane and the density profiles at 5 R$_\odot$. The results of the COCONUT simulations are also compared to tomography data. Different heating profiles generated different coronal configurations. The uniform spherically symmetric heating (introduced via Eq.~\ref{exp_heating}) did not obtain a realistic bi-modal solar wind configuration; however, the obtained solar wind at the outer boundary was fast. The heating profiles, including the magnetic field dependence (i.e. Eqs.~\ref{magnetic_damping} and~\ref{heating_lionello}) produced a bi-modal solar wind structure. When introducing heating via Eq.~\ref{heating_lionello}, the simulated density field was more similar to the one obtained in the MAS model. The next validation mechanism is to connect COCONUT output to the heliosphere and see how these effects propagate to Earth.
 
The bi-modal wind obtained near the outer coronal boundary is used as the onset of the heliospheric model in Icarus. COCONUT and Icarus are coupled by interpolating the plasma variables at 21.5 R$_\odot$ from the unstructured grid in COCONUT to the uniform grid used in Icarus. The heliospheric simulation in Icarus is initiated with the coronal boundary file from COCONUT. The initial conditions were relaxed to obtain the steady wind in the Icarus domain that stretched from 21.5 R$_\odot$ to 432 R$_\odot$. Furthermore, the time series were extracted at Earth and compared to OMNI 1-min data in order to assess the modelling of the various plasma conditions near Earth. The modelled heliosphere is more consistent with the OMNI data, with the heating introduced in Eq.~\ref{magnetic_damping} compared to the other heating profile. The radial velocity values were more similar in the case of the heating introduced via Eq.~\ref{magnetic_damping}; however, the number density was still strongly overestimated. The peak that was present in the number density data before the arrival of the higher-speed stream is better estimated by the same heating profile simulation. The difference in the magnetic field values is small at 1~AU and no strong conclusion can be made on this basis.

A comprehensive examination of the modelled data allowed us to assess the implemented full MHD model and identify the strengths and weaknesses of the model. Different heating approximations have show that it is important to deposit the thermal energy in the correct place to obtain a realistic bi-modal wind. Prescribing the uniform heating with exponential damping solely resulted in  accelerated wind everywhere near $\sim$ 21.5 R$_\odot$, but not in the bi-modal structure. Accumulating the thermal energy in the strong magnetic field regions resulted in a more realistic image of the solar wind. Both functions introduced in Eq.~\ref{magnetic_damping} and Eq.~\ref{heating_lionello} produced bi-modal wind near the outer boundary of the corona. This is the first achievement in the project since, as before, we could only get uniform solar wind distribution. The next question still remains regarding which solar wind is more realistic. The answer is important from the modelling point of view, with the goal to obtain as accurate results as possible, but also from the physics point of view. The main difference between these two approximated functions is that the first one treats the magnetic field only depending on its strength and introduces heating energy depending on the magnitude of the field, whereas the second one takes into account the contribution from the quiet Sun heating and active region heating separately and approximates both regimes with different formulas. In the case of the eclipse of 2019 (considered in this paper), the heating function approximated by \cite{Lionello2009} produced more realistic results when compared to the tomography data. 

After comparing the different approximated heating profiles in the solar corona, we can see that we can get our first MHD results comparable to observations and the MAS model; however, crucial physics is still missing in our full MHD model. We intend to experiment with more physics-based heating profiles, including the gradient of density and magnetic field. We also intend to implement the wave turbulence-driven heating mechanism to model Alfvén waves and the heating associated with them. As a first step, we obtained the full MHD chain from the Sun to Earth that is efficient and capable of operating for space weather purposes. As demonstrated, the advanced techniques, such as a fully implicit solver and an unstructured grid, allow us to include complex physics in our equations without significantly slowing down the code. Furthermore, the propagation of the flux rope CMEs in the full MHD model ought to be considered in order to get a more realistic thermodynamic evolution of the flux rope compared to the polytropic model (\cite{Linan2023}, \cite{Guo2023}). Finally, we intend to extend our model and include more realistic physics phenomena in subsequent works. 

\begin{acknowledgements}
TB acknowledges the help and advice received from Jon linker and Cooper Downs while developing the full 3D MHD coronal model in COCONUT. This research has received funding from the European Union’s Horizon 2020 research and innovation programme under grant agreement No 870405 (EUHFORIA 2.0) and the ESA project "Heliospheric modelling techniques“ (Contract No. 4000133080/20/NL/CRS).
These results were also obtained in the framework of the projects C14/19/089  (C1 project Internal Funds KU Leuven), AFOSR FA9550-18-1-0093, G.0B58.23N and G.0025.23N  (FWO-Vlaanderen), SIDC Data Exploitation (ESA Prodex-12), and Belspo project B2/191/P1/SWiM.
The Computational resources and services used in this work were provided by the VSC-Flemish Supercomputer Center, funded by the Research Foundation Flanders (FWO) and the Flemish Government-Department EWI.
\end{acknowledgements}


\bibliographystyle{aa}
\bibliography{bibliography}

\end{document}